\documentclass[aps,prb,twocolumn,superscriptaddress]{revtex4-2}

\usepackage{amsmath}    

\usepackage{amsfonts}
\usepackage{amssymb}
\usepackage[colorlinks=true,citecolor=blue,linkcolor=red]{hyperref}
\usepackage{graphicx}
\usepackage{bbold}					
\usepackage[dvipsnames]{xcolor}

\usepackage{bm} 
\usepackage{dcolumn}
\usepackage{mathtools}

\usepackage{array,booktabs,ragged2e}	
\usepackage{soul}						
\usepackage{cancel}						
\usepackage{tabularx}
\usepackage{multirow}
\usepackage{hhline}
\usepackage{adjustbox}
\usepackage{sidecap}
\usepackage{threeparttable}
\usepackage{colortbl} 
\usepackage{tabularx}
\usepackage[utf8]{inputenc}
\usepackage[T1]{fontenc}

\def\bra#1{\left<{#1}\right|}					
\def\ket#1{\left|{#1}\right>}					

\newcolumntype{P}[1]{>{\raggedleft\arraybackslash}p{#1}}
\newcolumntype{R}[1]{>{\centering\arraybackslash}p{#1}}

\usepackage{bm,color,amsmath,amssymb,mathrsfs,latexsym,graphicx,psfrag,mathtools}
\usepackage{color}
\usepackage[dvipsnames]{xcolor}

\usepackage{empheq}

\newcommand{\bsl}[1]{\boldsymbol{#1}}

\renewcommand{\mod}{\,\mathrm{mod}\,}

\newcommand{\ii}{\mathrm{i}}

\newcommand{\dsZ}{\mathbb{Z}}

\newcommand{\dsR}{\mathbb{R}}

\newcommand{\dsC}{\mathbb{C}}
\newcommand{\Tr}{\mathop{\mathrm{Tr}}}
\renewcommand{\Re}{\mathop{\mathrm{Re}}}

\newcommand{\eqnref}[1]{Eq.\,\eqref{#1}}

\newcommand{\refcite}[1]{Ref.\,\cite{#1}}

\newcommand{\mat}[1]{\left(\begin{matrix}#1\end{matrix}\right)}

\newcommand{\eq}[1]{\begin{equation} #1 \end{equation}}

\newcommand{\eqa}[1]{\begin{align}\begin{split} #1 \end{split}\end{align}}

\let\oldAA\AA
\renewcommand{\AA}{\text{\normalfont\oldAA}}

\newcommand{\ie}{{\emph{i.e.}}}
\newcommand{\eg}{{\emph{e.g.}}}
\newcommand{\TR}{\mathcal{T}}

\newcommand{\W}{\mathcal{W}}

\renewcommand{\P}{\mathcal{P}}

\newcommand{\N}{\mathcal{N}}

\newcommand{\diag}{\text{diag}}

\newcommand{\Ch}{\text{Ch}}

\usepackage{comment}

\usepackage{cleveref}
\crefname{appendix}{App.\,}{Apps.\,}
\crefname{equation}{Eq.\,}{Eqs.\,}
\crefname{figure}{Fig.\,}{Figs.\,}
\crefname{table}{Tab.\,}{Tabs.\,}
\crefname{section}{Sec.\,}{Secs.\,}
\creflabelformat{appendix}{[#2#1#3]}

\newcommand{\norm}[1]{\left\lVert#1\right\rVert}

\newcommand{\appBone}{\cref{app:non-abelian_stokes}}

\newcommand{\appC}{\cref{app:WL_bound}}
\newcommand{\appD}{\cref{app:Z2_bound}}
\newcommand{\appEone}{\cref{app:SW_bound}}
\newcommand{\appEtwo}{\cref{app:Optical_bound}}

\renewcommand{\appBone}{App.\,[B1]~\cite{SM}}

\renewcommand{\appC}{App.\,[C]~\cite{SM}}
\renewcommand{\appD}{App.\,[D]~\cite{SM}}
\renewcommand{\appEone}{App.\,[E1]~\cite{SM}}
\renewcommand{\appEtwo}{App.\,[E2]~\cite{SM}}

\begin{document}

\title{Universal Wilson loop Bound of Quantum Geometry}

\author{Jiabin Yu}
\email{yujiabin@ufl.edu}
\affiliation{Department of Physics, University of Florida, Gainesville, FL, USA}
\affiliation{Department of Physics, Princeton University, Princeton, New Jersey 08544, USA}

\author{Jonah Herzog-Arbeitman}
\affiliation{Department of Physics, Princeton University, Princeton, New Jersey 08544, USA}

\author{B.~Andrei Bernevig}
\affiliation{Department of Physics, Princeton University, Princeton, New Jersey 08544, USA}
\affiliation{Donostia International Physics Center, P. Manuel de Lardizabal 4, 20018 Donostia-San Sebastian, Spain}
\affiliation{IKERBASQUE, Basque Foundation for Science, Maria Diaz de Haro 3, 48013 Bilbao, Spain}

\begin{abstract}
We define the absolute Wilson loop winding and prove that it bounds the (integrated) quantum metric from below. This Wilson loop lower bound naturally reproduces the known Chern and Euler bounds of the integrated quantum metric, and provides an explicit lower bound of the integrated quantum metric due to the time-reversal protected $\dsZ_2$ index, answering a hitherto open question. In general, the Wilson loop lower bound can be applied to any other topological invariants characterized by Wilson loop winding, such as the particle-hole $\dsZ_2$ index.
As physical consequences of the $\dsZ_2$ bound, we show that the time-reversal $\dsZ_2$ index bounds superfluid weight and optical conductivity from below, and bounds the direct gap of a band insulator from above.
\end{abstract}

\maketitle

\section{Introduction}

Quantum geometry is a fundamental framework for the geometric properties of Bloch wavefunctions, which underpins a wide range of physical phenomena.
A key quantity that characterizes quantum geometry is the quantum metric~\cite{Provost1980FSMetric,Fubini1904,Study1905,Martin.Souza.1999}, 
\eq{
\left[g_{\bsl{k}}\right]_{ij} = \frac{1}{2}\Tr\left[ \partial_{k_i} P_{\bsl{k}} \, \partial_{k_j} P_{\bsl{k}} \right]\ ,
}
where $i,j$ denote spatial components, $\bsl{k}$ is Bloch momentum, and $P_{\bsl{k}}$ is the projector for the isolated set of bands of interest.
The quantum metric plays a pivotal role in diverse physics, including optical conductivity~\cite{SWM2000,Resta2006Polarization,Martin2004ElectronicStructure,Noack.Aebischer.2001,Queiroz.Verma.2024.instantaneous,Fu.Onishi.2024.Dielectric,Stengel.Souza.2024}, the superfluid weight of superconductors~\cite{Torma2015SWBoundChern,Torma2016SuperfluidWeightLieb,Liang2017SWBandGeo,Hu2019MATBGSW,Xie2020TopologyBoundSCTBG,Torma2020SFWTBG,Rossi2021CurrentOpinion,Yu2022EOCPTBG,Torma2023WhereCanQuantumGeometryLeadUs,Tian2023QuantumGeoSC}, electron-phonon coupling~\cite{Yu05032023GeometryEPC,Alexandradinata2024ShiftCurrent},  correlated charge fluctuations~\cite{Yu2024_QG_Charge_Fluctuation,Wu2024QGCornerChargeFluctuationsManyBody}, and fractional Chern insulators~\cite{neupert, sheng, regnault,Sun2011,Tang11}.
(See \refcite{Yu2025QGReview} for a review.)

In particular, many of the key roles played by the quantum metric in physical phenomena stem from the fact that its integral can be bounded from below by topological invariants.
For example, in superconducting twisted bilayer graphene~\cite{Cao2018TBGSC}, phase coherence (\ie, the nonzero superfluid weight) in topological flat bands is guaranteed by a nonzero Euler number~\cite{Ahn2018MonopoleNLSM,PhysRevLett.123.036401,Ahn2019TBGFragile}, which provides a lower bound on the quantum metric and, consequently, the geometric contribution to the superfluid weight~\cite{Xie2020TopologyBoundSCTBG,Yu2022EOCPTBG,Song20211110MATBGHF}.
Another example is that when the integrated quantum metric saturates the lower bound imposed by the Chern number~\cite{TKNN}, it becomes possible to construct analytical wavefunctions for fractional Chern insulators in continuum models~\cite{Jie2021IdealBands,Parker2023IdealBands,Valentin2023IdealBands,liu2024theorygeneralizedlandaulevels,Roy2014QGChernBound, claassen2015position, northe2022interplay,ji2024quantum,Roy2024FCINonLL}.
Despite the fundamental physical importance of the lower bound on the integrated quantum metric, explicit bounds have been established only in a limited set of cases—namely for the Chern number~\cite{Bellissard1994QGChernBound,Roy2014QGChernBound}, Euler number~\cite{Xie2020TopologyBoundSCTBG,Yu2022EOCPTBG,BJY2024EulerBoundQG,Slager2024EulerOptical}, chiral winding number~\cite{Torma2016QGChiralWinding}, and certain special two-dimensional obstructed atomic limits and fragile topological states~\cite{Herzog-Arbeitman2021QGOAI}—falling far short of encompassing the full range of known topological classifications in quantum materials~\cite{Kitaev2009TenFoldWayTITSC,Ryu2010TenFoldWayTITSC,Teo2010TenFoldWayTITSC,Bradlyn2017TQC,Po2020SI}.
Establishing explicit lower bounds for other topological invariants would enable new and deeper physical insights into all phenomena involving the quantum metric---an example of fundamental importance is the lower bound for the integrated quantum metric due to the Kane–Mele time-reversal ($\dsZ_2$) invariant~\cite{Kane2005Z2,Zhang2006QSH,Kane2005QSH,Bernevig2006BHZ}, which has remained a long-standing open question, despite that the $\dsZ_2$ index has been experimentally seen in various different systems~\cite{Konig2007QSHHgTe,Wu2018QSH,Kang2024_tMoTe2_2.13}.
Such new bounds may help us understand superconductivity in topological flat bands beyond twisted bilayer graphene (such as twisted bilayer MoTe$_2$~\cite{Xu_2025_SC_MoTe2,cai2023signatures,zeng2023integer,park2023observation,Xu2023FCItMoTe2,Ji2024LocalProbetMoTe2,Young2024MagtMoTe2,Kang2024_tMoTe2_2.13,xu2024interplaytopologycorrelationssecond,park_Ferromagnetism_2024,Park_2025_tMoTe2_gap}), as well as topologically ordered phases beyond fractional Chern insulators (\eg, fractional topological insulators~\cite{bernevig2006quantum2,Levin_Stern,Neupert2011FTI,Stern2015review,Neupert_2015,Levin2012classification}).
In this context, we emphasize that the central goal is to obtain explicit expressions for the lower bounds imposed by given topological invariants, as these are essential for making physically meaningful predictions. 

In this work, we present an explicit expression for the lower bound of the integrated quantum metric arising from any form of band topology—whether stable, fragile, or obstructed atomic—that is characterizable by Wilson loop (WL) winding.
WL winding is known for capturing a wide range of different band topology, encompassing the Chern number~\cite{BAB2014WLInversion}, Euler number~\cite{Ahn2018MonopoleNLSM,PhysRevLett.123.036401,Ahn2019TBGFragile}, time-reversal ($\mathbb{Z}_2$) index~\cite{Dai2011Z2WilsonLoop}, and various other topological invariants~\cite{PhysRevB.100.195135,PhysRevB.99.045140}.
As a corollary, our result naturally resolves the long-standing open question of deriving an explicit lower bound on the integrated quantum metric imposed by the Kane–Mele $\mathbb{Z}_2$ invariant.
Specifically, we define the ``absolute WL winding'', which is the sum of the absolute phase changes of all WL eigenvalues without cancellation.
Then, with the non-abelian Stokes theorem~\cite{Arefeva1980nonabelianstokes}, we can prove that the integrated $\Tr[g_{\bsl{k}}]$ and $2\sqrt{\det(g_{\bsl{k}})}$ is bounded from below by the absolute WL winding.
\eq{
\label{main_eq:QG_WL_bound}
\int_{BZ} d^2 k\ \Tr[g_{\bsl{k}}]  \geq \int_{\text{BZ}} d^2 k\ 2\sqrt{\det(g_{\bsl{k}})} \geq \overline{\N}\ ,
}
where $\overline{\N}$ is the absolute WL winding for any proper deformation that ends up covering the entire BZ, and the integration is over the Brillouin zone (BZ).
As a consequence, we derive the TR-$\dsZ_2$ lower bound of the integrated quantum metric in two spatial dimensions, owing to $\overline{\N} \geq 4\pi \dsZ_2$ for the choice of the deformation that is approprate for $\dsZ_2$.
The $\dsZ_2$ bound also holds for the particle-hole (PH) protected $\dsZ_2$ index, which, as \refcite{Song2020TBGII} points out, is equivalent to the TR-$\dsZ_2$ index.
We further describe how the general inequality \cref{main_eq:QG_WL_bound} provides lower bounds for superfluid weight and optical conductivity and an upper bound for the direct gap of a band insulator.

\section{General Wilson Loop Bound of the Integrated Quantum Metric}

\begin{figure}[t]
    \centering
    \includegraphics[width=\linewidth]{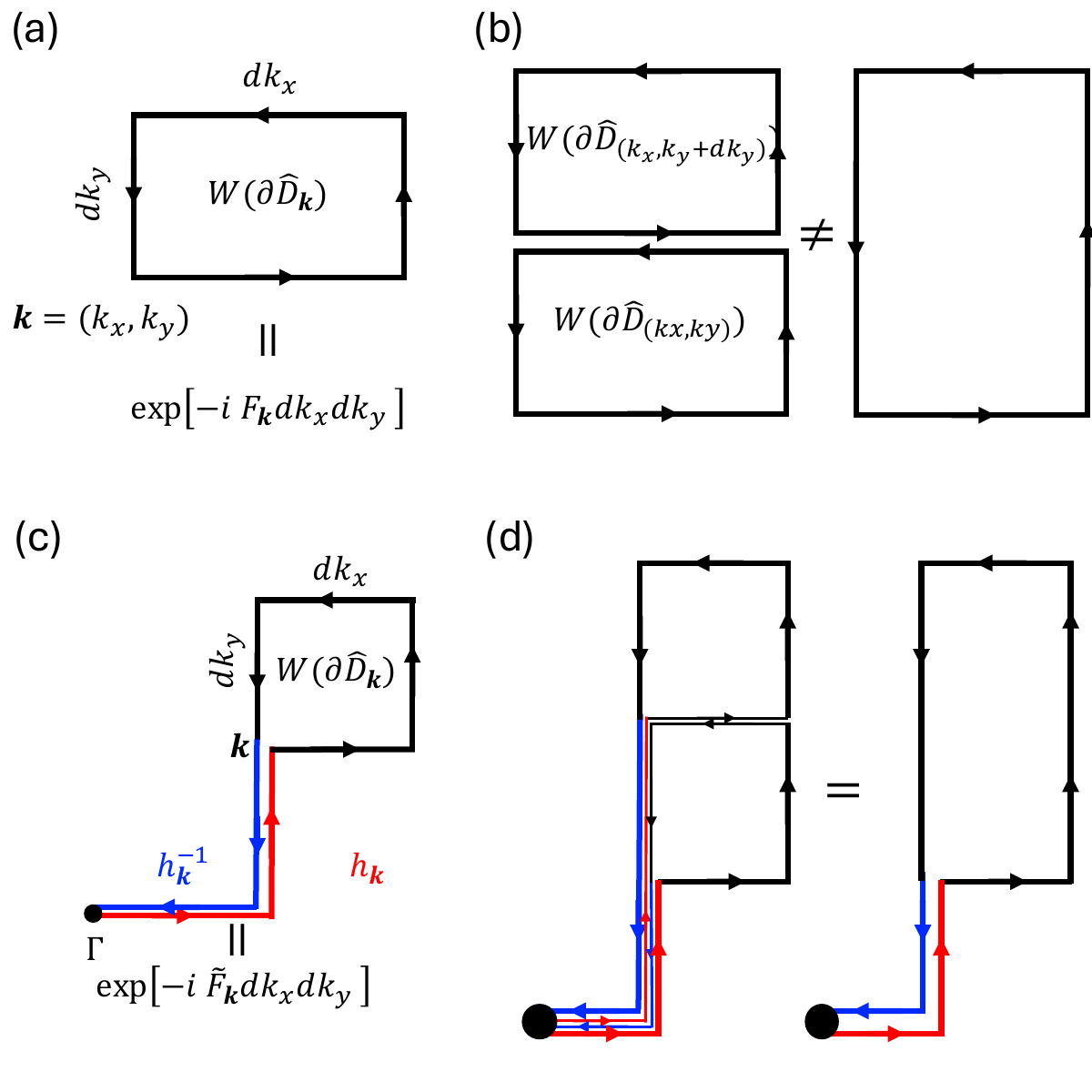}
    \caption{(a) shows the WL of an infinitesimal $\hat{D}_{\bsl{k}} = [k_x,k_x +dk_x] \times [k_y,k_y +dk_y]$, which equals $\exp\left[ - \ii  F_{\bsl{k}} dk_x dk_y\right] $ with $F(\bsl{k})$ the non-abelian Berry curvature.
    (b) shows that the multiplication of WLs $W\left( \partial \hat{D}_{(k_x,k_y)} \right)$ and $W\left( \partial  \hat{D}_{(k_x,k_y+dk_y)} \right)$ is not necessarily $W\left( \partial\left[ \hat{D}_{(k_x,k_y)}\cup \hat{D}_{(k_x,k_y+dk_y)} \right]\right) $ owing to the non-abelian nature of the WL.
    (c) shows the WL of an infinitesimal $\hat{D}_{\bsl{k}} = [k_x,k_x +dk_x] \times [k_y,k_y +dk_y]$ combined with with dressing $h_{\bsl{k}}$ and $h_{\bsl{k}}^{-1}$, where $h_{\bsl{k}}$ is the red Wilson line, and the inverse $h_{\bsl{k}}^{-1}$ is the blue Wilson line.
    Here $\widetilde{F}(\bsl{k}) = h_{\bsl{k}} F_{\bsl{k}} h_{\bsl{k}}^{-1}$, the dressed non-abelian Berry curvature.
    (d) shows the merging of two dressed WLs into one dressed WL on a larger loop.  
    }
    \label{fig:non-abelian_stockes}
\end{figure}

We start by sketching the proof of the general WL lower bound of the integrated quantum metric.
The setup of the problem is an isolated set of $N$ bands of a 2D noninteracting lattice system, and we label the periodic parts of the relevant Bloch states as $\ket{u_{n,\bsl{k}}}$ with $n=1,2,...,N$.
The general Wilson line for the isolated set of bands along path $\gamma$ is defined as~\cite{Dai2011Z2WilsonLoop} 
\eq{
\label{mian_eq:Wilson_line}
W(\gamma) = \lim_{L\rightarrow +\infty}\bra{u_{\bsl{k}_0}} P_{\bsl{k}_1} P_{\bsl{k}_2} \cdots P_{\bsl{k}_{L-1}}  P_{\bsl{k}_{L}} \ket{u_{\bsl{k}_L}} \ ,
}
where $P_{\bsl{k}}=\sum_n \ket{u_{n,\bsl{k}}} \bra{u_{n,\bsl{k}}}$, $\bsl{k}_0$, $\bsl{k}_1$, $\bsl{k}_2$, ..., $\bsl{k}_L$ are sequential momenta on $\gamma$ with $\bsl{k}_0$ at the beginning and $\bsl{k}_L$ at the end.
\cref{mian_eq:Wilson_line} becomes a WL when $\gamma$ forms a loop.
Meanwhile, the non-abelian Berry curvature for the isolated set of bands reads
\eq{
F_{\bsl{k}} = \partial_{k_x} A_{\bsl{k},y} - \partial_{k_y}A_{\bsl{k},x} - \ii [A_{\bsl{k},x},A_{\bsl{k},y}]\ ,
}
where
\eq{
\left[ A_{\bsl{k},i} \right]_{mn} = \ii  \bra{u_{m,\bsl{k}}} \partial_{k_i}\ket{u_{n,\bsl{k}}}\ .
}

The first step of the proof is to relate the WL to the non-abelian Berry curvature, which can be done through a known theorem called non-abelian Stokes theorem~\cite{Arefeva1980nonabelianstokes}.
The theorem holds for any simply connected region $D$; but for simplicity, we, in the main text, consider a rectangular region  $D= [0,K_x]\times [ 0,K_y]$, and we label the boundary (counter-clockwise) of $D$ as $\partial D$.
For the WL along $\partial D$ (labeled as $W(\partial D)$ according to \cref{mian_eq:Wilson_line}) with starting momentum $\bsl{k}_0=(0,0)$, the non-abelian Stokes theorem states that
\eqa{
\label{mian_eq:non-abelian_stokes}
& W(\partial D) = \mathcal{P} \exp\left[ - \ii \int_D \widetilde{F}_{\bsl{k}} dk_x dk_y\right]   \\
&=  \prod_{i=L_1-1,...,1,0}  \prod_{j=0,1,...,L_2-1}  \exp\left[ - \ii  \widetilde{F}_{(\frac{ i K_x}{L_1}, \frac{j K_y}{L_2})} \frac{K_x}{L_1}  \frac{K_y}{L_2}\right] \ ,
}
where $L_1$ and $L_2$ limit to infinity, $\mathcal{P}$ is the path ordering that moves larger $k_x$ and smaller $k_y$ to the left as shown by the second equality.
$\widetilde{F}_{\bsl{k}}$ is the dressed non-abelian Berry curvature
\eq{
\label{main_eq:dressed_F}
\widetilde{F}_{\bsl{k}} = h_{\bsl{k}} F_{\bsl{k}} h_{\bsl{k}}^{-1}\ ,
}
where $h_{\bsl{k}} = W_{\bsl{k}_0\rightarrow (k_x,0)}  W_{(k_x0)\rightarrow \bsl{k}}$ with $\bsl{k}_0\rightarrow \bsl{k}_1$ denoting the straight path from $\bsl{k}_0$ to $\bsl{k}_1$.

We now provide the intuitive reason  why \cref{mian_eq:non-abelian_stokes} uses the dressed non-abelian Berry curvature $\widetilde{F}_{\bsl{k}}$ instead of the non-abelian Berry curvature $F_{\bsl{k}}$. A detailed proof of \cref{mian_eq:non-abelian_stokes} is presented in {\appBone}.
Let us first note that the WL along the boundary of an infinitesimal plaquette is just the exponential of the non-abelian Berry curvature (to leading order in $dk_x dk_y$), $\exp\left[ - \ii  F_{(k_x,k_y)} dk_x dk_y\right] $, as shown in \cref{fig:non-abelian_stockes}(a).
If we directly multiply the WLs of two neighboring infinitesimal plaquettes, the result does not necessarily equal the WL along the boundary of the union of the two plaquettes, owing to its non-abelian nature.
This means that if we multiply $\exp\left[ - \ii  F_{(k_x,k_y)} dk_x dk_y\right] $ for all plaquettes in $D$, no matter how we select the path, we cannot necessarily arrive at the WL $W_{\partial D} $.
The Wilson line dressing in \cref{main_eq:dressed_F} resolve this issue.
With the dressing,  multiplying $\exp\left[ - \ii  \widetilde{F}_{(k_x,k_y)} dk_x dk_y\right]$ for two nearby plaquettes  directly yields the dressed WL along boundary of the union of two plaquettes, as shown in \cref{fig:non-abelian_stockes}(c,d).
As we multiply the $\exp\left[ - \ii  \widetilde{F}_{(k_x,k_y)} dk_x dk_y\right]$ for all plaquettes in $D$, the dressing eventually cancels, and leads to \cref{mian_eq:non-abelian_stokes}.
We note that the appearance of the dressing is only needed when dealing with the nonabelian  (essentially multi-band) case such as $\dsZ_2$ index---it is not necessary in the abelian case (such as Chern number or the Euler number in the Chern gauge).

\begin{figure}[t]
    \centering
    \includegraphics[width=\linewidth]{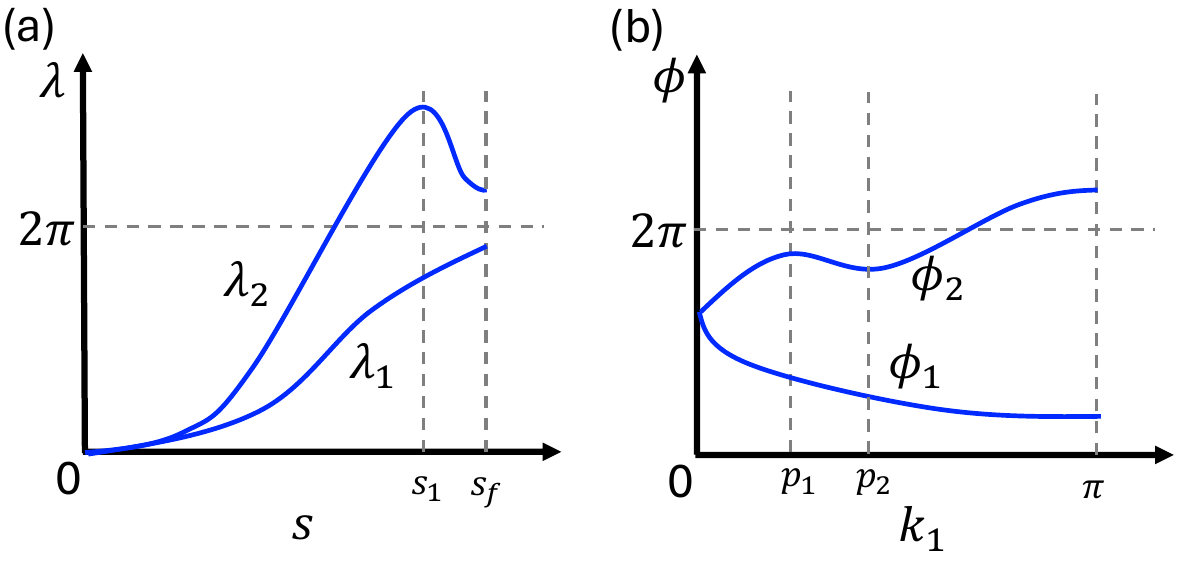}
    \caption{(a) is a schematic plot of a 2-band WL $W(\partial D_s)$, where $D_s$ is a disk-like region that continuously depends on $s \in [0,s_f]$ with $D_0$ an measure-zero region (such as a point). 
    The absolute WL winding of $W(\partial D_s)$ is equal to $|\lambda_{2,s_1}-\lambda_{2,0}|+|\lambda_{2,s_f}-\lambda_{2.s_1}| +|\lambda_{1,s_f}-\lambda_{1,0}| $.
    (b) is a schematic plot of a 2-band WL $\mathcal{W}(k_1)$ in \cref{main_eq:WL_Z2} for $\dsZ_2 = 1$. 
    The absolute WL winding of $\mathcal{W}(k_1)$ is equal to $|\phi_2(p_1)-\phi_2(0)|+|\phi_2(p_2)-\phi_2(p_1)|+|\phi_2(\pi)-\phi_2(p_2)| +|\phi_1(\pi)-\phi_1(0)| $, which is no smaller than $\phi_2(\pi)-\phi_1(\pi) = 2\pi \dsZ_2$.
    }
    \label{fig:WL_winding}
\end{figure}

With the non-abelian Stokes theorem, we now consider the region $D_s$ that continuously and monotonically depends on $s \in [0,s_f]$ with $D_0$ having zero area.
Here by ``monotonically'', we require that $D_s\subset D_{s'}$ for any $s\leq s'$.
We refer to this deformation as a proper deformation.
The WL $W(\partial D_s)$ now depends continuously on $s$, and we require the starting point $\bsl{k}_0$ of $W(\partial D_s)$ is the same for all $s\in[0,s_f]$.
$W(\partial D_s)$ has eigenvalues $e^{\ii \lambda_{i,s} }$ with $i=1,2,...,N$, and without loss of generality, we will always choose $\lambda_{i,s}$ to continuously depend on $s$. 
We can then prove (in {\appC}) that
\eqa{
\label{main_eq:WL_lower_bound_of_rho_Ftilde}
\int_0^{s_f} ds \sum_{i} \left|\frac{d\lambda_{i,s}}{ds} \right| & \leq \int_0^{s_f} ds \ \rho\left[ -\ii   \partial_{s} \log W(\partial D_s) \right] \\
& \leq \int_{D_{s_f}} d^2k\ \rho(F_{\bsl{k}}) \ ,
}
where $\rho(A)$ is the sum of the absolute values of all eigenvalues of the matrix A, which is called the Schatten $1$-norm~\cite{bhatia2009positive}, and we have used the fact that $\widetilde{F}_{\bsl{k}}$ and $F_{\bsl{k}}$ have the same eigenvalues.
The proof of \eqnref{main_eq:WL_lower_bound_of_rho_Ftilde} exploits the properties (especially the triangle inequality) of the Schatten $1$-norm, which is elaborated in {\appC}.
Since $\rho(F_{\bsl{k}}) \leq 2\sqrt{\det(g_{\bsl{k}})} \leq \Tr[g_{\bsl{k}}]$ as shown in {\appC}, we arrive at the WL lower bound of the integrated quantum metric
\eqa{
\label{main_eq:WL_lower_bound_of_Trg}
 \int_0^{s_f} ds \sum_{i} \left|\frac{d\lambda_{i,s}}{ds} \right| \leq \int_{D_{s_f}}  d^2 k 2\sqrt{\det(g_{\bsl{k}})} \leq \int_{D_{s_f}}  d^2 k \Tr[g_{\bsl{k}}] \ .
}
We refer to $ \int_0^{s_f} ds \sum_{i} \left|\frac{d\lambda_{i,s}}{ds} \right|$ in \cref{main_eq:WL_lower_bound_of_Trg} as the ``absolute WL winding'', because it counts the winding of the WL eigenvalues without cancellation, as shown in \cref{fig:WL_winding}(a).
We would recover \cref{main_eq:QG_WL_bound} as long as $D_{s_f}=\text{BZ}$.
We emphasize that the absolute WL winding does not depend on how we rank $\lambda_{i,s}$ as long as they are continuous; it is because $\lambda_{i,s}$ generally touch or cross at measure-zero points of $s$ which can be neglected for the integral.
Naturally, the absolute WL winding is no smaller than the absolute value of the total WL winding, which is $\left|\int_0^{s_f} ds \sum_{i} \frac{d\lambda_{i,s}}{ds} \right|$.

More importantly, there is an freedom in defining the absolute WL winding while keeping \cref{main_eq:WL_lower_bound_of_Trg} valid.
Let us define $\mathcal{W}_{s}$ such that
\eq{
\label{main_eq:W_s}
W(\partial D_s) =  U_s \mathcal{W}_{s} U_s^\dagger   V
}
with unitary $V$ and $U_s$ and $U_s$ depending on $s$ smoothly, and define
$\phi_{l}(s)$ to be the phase of the $l$th eigenvalue of $\mathcal{W}_{s}$.
As long as we choose $\phi_{l}(s)$ to be continuous, we have
\eqa{
\label{main_eq:WL_lower_bound_of_Trg_for_Z_2}
\mathcal{N} & \equiv \int_0^{s_f} ds \sum_{l} \left|\frac{d\phi_{l}(s)}{ds} \right| \leq \int_{D_{s_f}}  d^2 k 2\sqrt{\det(g_{\bsl{k}})} \\
& \leq \int_{D_{s_f}}  d^2 k \Tr[g_{\bsl{k}}] \ ,
}
where $\mathcal{N}$ is called the absolute WL winding of the proper deformation $D_s$ and dressing $V$ and $U_s$.
\cref{main_eq:WL_lower_bound_of_Trg_for_Z_2} comes from the fact that $V$ is independent of $s$ and $ U_s \W_s U_s^\dagger$ have the same eigenvalues as $\W_s$, which gives
\eqa{
& \rho\left[ -\ii \partial_{s} \log W(\partial D_s) \right]   = \rho\left[ -\ii \partial_{s} \log   U_s \W_s U_s^\dagger  \right]\\
& \geq \int_0^{s_f} ds \sum_{l} \left|\frac{d\phi_{l}(s)}{ds} \right|\ .
}
Combined with \cref{main_eq:WL_lower_bound_of_Trg}, we arrive at \cref{main_eq:WL_lower_bound_of_Trg_for_Z_2}.

Of particular usefulness (\eg, for the derivation of the $\dsZ_2$ bound), is the case when $D_{s} = \{ k_1 \bsl{b}_1/(2\pi) + k_2 \bsl{b}_2/(2\pi) | k_1\in[0,s]\ , k_2\in[0,2\pi] \}$.
Here $\bsl{b}_1$ and $\bsl{b}_2$ are two basis reciprocal lattice vectors, which means $D_{2\pi}=\text{BZ}$ is the entire first Brillouin zone.
In this case, $W\left(\partial D_{s}\right)$ has the same expression in \cref{main_eq:W_s} with $U_s = W\left(\Gamma \rightarrow s \bsl{b}_1/ (2\pi)\right) $, $V= W\left(  \bsl{b}_2 \rightarrow \Gamma\right)$, and
\eq{
\label{main_eq:WL_Z2}
\mathcal{W}_{s} = W\left( s \bsl{b}_1/ (2\pi) \rightarrow s \bsl{b}_1/ (2\pi) + \bsl{b}_2 \right)\ ,
}
which is nothing but a new WL along $\bsl{b}_2$ at a fixed fraction of $\bsl{b}_1$. ($\Gamma = (0,0)$).
With this choice, we can directly derive the known Chern bound~\cite{Roy2014QGChernBound} by choosing $s_f = 2\pi$.
In this case, the total WL winding of $\mathcal{W}_s$ is just $2\pi\Ch$ (with $\Ch$ the Chern number), which directly gives the Chern bound
$|\Ch| \leq \frac{1}{2\pi} \int_{\text{BZ}}  2 \sqrt{\det(g_{\bsl{k}})}  dk_x dk_y\leq \frac{1}{2\pi} \int_{\text{BZ}}  \Tr[g_{\bsl{k}}]  dk_x dk_y$.
For Euler number $e_2 \in \mathbb{Z}$ defined for an isolated set of two bands with combination of time-reversal and inversion (or two-fold rotational) symmetries, the absolute WL winding would be no smaller than $4\pi |e_2|$, as we can always choose a Chern gauge where we can view the two components of the basis as two Chern bands with Chern numbers $\pm e_2$.
Therefore, we can reproduce the known Euler bound~\cite{Xie2020TopologyBoundSCTBG,Yu2022EOCPTBG,BJY2024EulerBoundQG,Slager2024EulerOptical}, which is $2 |e_2| \leq \frac{1}{2\pi} \int_{\text{BZ}}  2 \sqrt{\det(g_{\bsl{k}})}  dk_x dk_y \leq \frac{1}{2\pi} \int_{\text{BZ}}  \Tr[g_{\bsl{k}}]  dk_x dk_y$.

\section{$\dsZ_2$ Bound of the Integrated Quantum Metric}

Besides reproducing the known Chern and Euler bounds, \cref{main_eq:WL_lower_bound_of_Trg} can also provide the TR-protected $\dsZ_2$ bound of the integrated quantum metric, which has remained an open question up to now.

To prove the bound, let us first review how to calculate TR-protected $\dsZ_2$ index from WL winding.
The setup is an isolated set of $2N$ bands of a non-interacting system with non-negligible spin-orbit coupling. The number of bands is necessarily even due to Kramers degeneracy of spinful TR symmetry $\mathcal{T}$.
To calculate the $\dsZ_2$ index of the isolated set of bands, we will use the WL $\mathcal{W}_{k_1}$ in \cref{main_eq:WL_Z2}, with eigenvalues $e^{\ii \phi_{l}(k_1)}$ with $l=1,2,...,2N$.
Without loss of generality, we can choose $\phi_{l}(k_1)$ to be continuous for $k_1\in [0,\pi]$ (see \cref{fig:WL_winding}(b) for an example and see {\appD} for details). We can fix $\phi_l(0)\in [0,2\pi)$.
In this case, $\dsZ_2$ reads \cite{Dai2011Z2WilsonLoop}
\eq{
\label{main_eq:Z2_Dai}
\dsZ_2 = \sum_{l=1}^{2N} M_l\ \mod\ 2 \ ,
}
where $2 \pi M_l = \phi_l(\pi) - \left[\phi_l(\pi)\ \mod\ 2\pi \right] $ with $\left[ x\ \mod\ 2\pi \right]\in [0,2\pi)$ and $M_l\in\dsZ$.

Now we discuss how $\dsZ_2$ bounds the quantum geometry from below.
As exemplified in \cref{fig:WL_winding}(b) and elaborated in {\appD}, the $\dsZ_2$ index in \cref{main_eq:Z2_Dai} suggests that the absolute WL winding of  $\mathcal{W}_{k_1}$ over $k_1\in [0,\pi]$ must be no smaller than $2\pi \dsZ_2$, \ie,
\eq{
\label{main_eq:absolute_winding_Z_2}
\frac{1}{2\pi}\int_0^{\pi} dk_1 \sum_{l} \left|\frac{d\phi_{l}(k_1)}{d k_1} \right| \geq  \dsZ_2\ ,
}
where $\phi_{l}(k_1)$ is the phase of the $l$th eigenvalue of $\mathcal{W}_{k_1}$, and we choose $\phi_{l}(k_1)$ to be continuous. 
Combined with \cref{main_eq:WL_lower_bound_of_Trg_for_Z_2} and TR symmetry, we arrive at the inequality 
\eq{
\label{main_eq:QG_Z_2_bound}
\frac{1}{2\pi}\int_{BZ} d^2 k\ \Tr[g_{\bsl{k}}]  \geq \frac{1}{2\pi} \int_{\text{BZ}} d^2 k\ 2\sqrt{\det(g_{\bsl{k}})} \geq 2 \dsZ_2\ .
}
This inequality can be saturated, as two TR-related lowest Landau levels with opposite spins serve as one example.

\cref{main_eq:QG_Z_2_bound} also holds for the PH-protected $\dsZ_2$ index, where the $PH$ symmetry is antiuniatry and squares to $-1$.
It comes from the equivalence between the TR-$\dsZ_2$ and PH-$\dsZ_2$ indices, which was pointed out in \cite{Song2020TBGII}, as reviewed in {\appD}.
We note that \cref{main_eq:QG_Z_2_bound} also holds if we replace $\sqrt{\det(g_{\bsl{k}})}$ by the larger diagonal element of the quantum metric. (See details in {\appD}.)

\section{Physical Consequences of the Wilson-Loop Bound}

The WL bound (and thsu the $\dsZ_2$ bound) can provide bounds on various physical quantities.
The first example is the superfluid weight of 2D superconductors.
Consider a set of isolated two normal-state bands that preserve TR symmetry. Assuming a mean-field treatment in the flat-band limit where the pairing potential $\Delta$ is uniform over the BZ (see {\appEone}), the superfluid weight is proportional to the minimal integrated quantum metric \cite{Torma2015SWBoundChern,Torma2022ReviewQuantumGeometry,Huhtinen2022FlatBandSCQuantumMetric}. Hence the WL bound yields
\eq{
\Tr[D_{SW}(T=0)] \geq \frac{4 |\Delta|}{\pi}  \sqrt{f (1-f)} \frac{\overline{\N}}{4\pi^2} \ ,
}
where $\overline{\N}$ is the absolute WL winding of any proper deformation that ends up covering the whole BZ (and of any dressing), $D_{SW}(T=0)$ is the zero-temperature superfluid weight tensor, $|\Delta|$ is the strength of the pairing, and $f\in[0,1]$ is the occupation fraction of the two bands ($f=1$ means the two bands are fully filled).
Here and henceforth we take $\hbar=e=1$ with electron having charge $-e$.
The WL bound of the superfluid weight simply leads to the $\dsZ_2$ bound of the superfluid weight due to \cref{main_eq:absolute_winding_Z_2}.
As a result of this $\dsZ_2$ bound, a spin-orbital-coupled flat band system will definitely have phase coherence after the formation of uniform Cooper pair, as long as it has nontrivial $\dsZ_2$ (or any nontrivial topology characterized by WL).
Recently, superconductivity has been observed in twisted bilayer MoTe$_2$ at  partial filling of the two flat bands which possess nonzero $\dsZ_2$ index~\cite{Xu_2025_SC_MoTe2}; in this case, the phase coherence or nonzero superfluid weight may be understood from the $\dsZ_2$-bounded integrated quantum metric, if the intervalley pairing dominates.

Besides superfluid weight, the optical conductivity of a band insulator has been shown to be related to its quantum geometry~\cite{SWM2000}
\eq{
\int_0^\infty \frac{d\omega}{\omega} \sum_i \Re[ \sigma_{ii}(\omega + \ii 0^+)] =    \frac{1}{4\pi} \int d^2k\ \Tr[g(\bsl{k})]\ ,
}
where $\sigma_{ij}(\omega + \ii 0^+)$ is the optical conductivity tensor, and $0^+$ is an infinitesimal real positive number.
Combined with the sum rule derived from the Kubo formula~\cite{Kubo1957}, it has been shown that the direct band gap $E_g$ is bounded above by the integrated quantum metric~\cite{Kivelson1982,Fu.Onishi.2023}
\eq{
E_g \leq\frac{ 2n \pi}{m } \frac{1}{\frac{1}{4\pi} \int d^2k\ \Tr[g(\bsl{k})]} \ ,
}
where $n$ is the electron number density, and $m$ is the electron mass.
Combining these known results with the WL bound in \cref{main_eq:QG_WL_bound}, we can immediately arrive at the lower bound of integrated optical conductivity and the upper bound of the direct band gap, \ie,
\eqa{
& \int_0^\infty \frac{d\omega}{\omega} \sum_i \Re[ \sigma_{ii}(\omega + \ii 0^+)] \geq \frac{\overline{\N}}{4\pi}\\
& E_g \leq\frac{8 n \pi^2}{ m \overline{\N}} \ .
}
Again, these WL bounds can leads to a bound for the Kane-Mele invariant using $\overline{\N}\geq 4\pi \dsZ_2$ derived from \cref{main_eq:absolute_winding_Z_2}.
The upper bound intuitively aligns with the empirical observation that large-gap topological insulators are rare and difficult to realize.
More details are given in {\appEtwo}.

\section{Conclusion}

We show that the absolute WL winding bounds the integrated quantum metric from below, which provides the TR-$\dsZ_2$ and PH-$\dsZ_2$ lower bound of the integrated quantum metric, as well as recovering the previously known Chern and Euler bounds.
As a result, the TR-$\dsZ_2$ index provides lower bounds for superfluid weight and optical conductivity and an upper bound for the band gap of a band insulator.
The WL bound is applicable to any topological invariant that can be characterized by WL winding.
One interesting future direction is to generalize the bound to other geometric quantities.

\section{Acknowledgment}
J. Y.'s work at University of Florida is supported by startup funds at University of Florida.
J. Y.'s work at Princeton University is supported by the Gordon and Betty Moore Foundation through Grant No. GBMF8685 towards the Princeton theory program.
J. H.-A. is supported by a Hertz Fellowship, with additional support from DOE Grant No. DE-SC0016239.
B.A.B. was supported by the Gordon and Betty Moore Foundation through Grant No. GBMF8685 towards the Princeton theory program, the Gordon and Betty Moore Foundation’s EPiQS Initiative (Grant No. GBMF11070), the Office of Naval Research (ONR Grant No. N00014-20-1-2303), the Global Collaborative Network Grant at Princeton University, the Simons Investigator Grant No. 404513, the BSF Israel US foundation No. 2018226, the NSF-MERSEC (Grant No. MERSEC DMR 2011750), the Simons Collaboration on New Frontiers in Superconductivity, and the Schmidt Foundation at the Princeton University.


%

\appendix
\onecolumngrid

\tableofcontents

\section{Quantum Geometric Tensor, Quantum Metric, and Berry Curvature}
\label{app:QG_review}

In this appendix, we discuss the key concepts used in our work.

\subsection{Review}

We first review several key definitions following \refcite{Yu2025QGReview}.
Consider an isolated set of $N$ bands, and we label the periodic part of the Bloch states as $\ket{u_{\bsl{k},n}}$ with $n=1,2,...,N$.
The quantum geometric tensor for the set of bands reads
\eq{
\left[Q_{i j}(\bsl{k})\right]_{mn} = \bra{\partial_{i} u_{\bsl{k},m}} (1-P_{\bsl{k}}) \ket{\partial_{j} u_{\bsl{k},n}} \ ,
}
where $P_{\bsl{k}}=\sum_{n}\ket{u_{\bsl{k},n}}\bra{u_{\bsl{k},n}}$, and $\partial_{i} = \partial_{k_{i}}$.
The non-abelian Berry curvature reads
\eq{
\label{eq:F_def}
\left[F_{ij,\bsl{k}} \right]_{mn}= \ii \left[ Q_{i j}(\bsl{k}) - Q_{j i}(\bsl{k}) \right]_{mn}\ ,
}
and the non-abelian quantum metric reads
\eq{
\label{eq:G_def}
\left[ G_{ij,\bsl{k}} \right]_{mn}= \frac{1}{2}  \left[ Q_{i j}(\bsl{k}) + Q_{j i}(\bsl{k}) \right]_{mn} \ .
}
Both $F_{ij,\bsl{k}}$ and $G_{ij,\bsl{k}}$ are Hermitian matrix for fixed $i,j$:
\eq{
F_{ij,\bsl{k}} ^ \dagger = F_{ij,\bsl{k}} \ ,\ G_{ij,\bsl{k}} ^ \dagger = G_{ij,\bsl{k}} \ .
}
The quantum metric reads
\eqa{\label{eq:g}
\left[g_{\bsl{k}}\right]_{ij}= \Tr[G_{ij,\bsl{k}}] = \frac{1}{2}\Tr \left[ (\partial_{i} P_{\bsl{k}}) \partial_{j}P_{\bsl{k}} \right] \ .
}

\subsection{Bounds on Quantum Metric By Nonabelian Berry Curvature}

Now we discuss how the nonabelian Berry curvature bounds quantu metric from below.
$\left[Q_{i j}(\bsl{k})\right]_{mn}$ is positive semi-definite.
To see this, let us consider a generic $U_{im}$, which gives
\eq{
\sum_{ij,mn} \left[Q_{i j}(\bsl{k})\right]_{mn} U_{im}^* U_{jn} = \sum_{ij,mn}  U_{im}^* \bra{\partial_{i} u_{\bsl{k},m}} (1-P_{\bsl{k}}) \ket{\partial_{j} u_{\bsl{k},n}} U_{jn} = \bra{X}  (1-P_{\bsl{k}}) \ket{X}\ ,
}
where $\ket{X} = \sum_{jn}  \ket{\partial_{j} u_{\bsl{k},n}} U_{jn} $.
Since $1-P_{\bsl{k}}$ only has nonnegative eigenvalues, we know
\eq{
\label{eq:positive_semideifinte_Q}
\sum_{ij,mn} \left[Q_{i j}(\bsl{k})\right]_{mn} U_{im}^* U_{jn} \geq 0
}
for any $U_{jn}$, meaning that $\left[Q_{i j}(\bsl{k})\right]_{mn}$ is positive semi-definite.

The positive semi-definiteness of the quantum geometric tensor provides an inequality between quantum metric and the non-abelian Berry curvature.
To show it, let us choose $U_{im}$ in \cref{eq:positive_semideifinte_Q} as $U_{im} = (\delta_{i,i_0} + s \ii \delta_{i,j_0})v_m$ for $s=\pm 1$ and arbitrary $v_m$.
In this case, we have
\eqa{
& \sum_{mn} v_m^* \left[Q_{i_0 i_0}(\bsl{k}) + Q_{j_0 j_0}(\bsl{k}) + s \ii Q_{i_0 j_0}(\bsl{k})- s \ii Q_{j_0 i_0}(\bsl{k})\right]_{mn} v_n  \geq 0 \\
& \Rightarrow v^\dagger \left[ G_{i_0 i_0,\bsl{k}} + G_{j_0 j_0,\bsl{k}} + s F_{i_0 j_0,\bsl{k}}\right] v  \geq 0 
}
for any  $i_0$, $j_0$ and vector $v$.
In 2D, we define 
\eq{
\label{eq:F_xy_in_short_as_F}
F_{\bsl{k}} = F_{xy,\bsl{k}}\ ,
}
which gives
\eq{
v^\dagger \left[ \sum_i G_{ii,\bsl{k}}+ s F_{\bsl{k}}\right] v \geq 0 \Rightarrow  v^\dagger \left[ \sum_i G_{ii,\bsl{k}}\right] v \geq  \left| v^\dagger  F_{\bsl{k}} v \right|
}
for any vector $v$.
Eventually, we arrive at, in 2D,
\eq{
\label{eq:Trg_non-abelian_Berry}
\Tr[g_{\bsl{k}} ] \geq \sum_l \left| v_l^\dagger  F_{\bsl{k}} v_l \right|
}
for any choice of a complete set of orthonormal $v_l$. A special choice of $v_l$ is the eigenbasis of $F_{\bm{k}}$, resulting in 
\eq{
\Tr[g_{\bsl{k}} ] \geq  \rho(F_{\bsl{k}})\ ,
}
where $\rho(F_{\bsl{k}})$ is the sum of the absolute values of all eigenvalues of $F_{\bsl{k}}$.

Besides $\Tr[g_{\bsl{k}} ]$, $\rho(F_{\bsl{k}})$ can also bound $\sqrt{\det(g_{\bsl{k}})}$ from below in 2D.
Let us choose $U_{im}$ in \cref{eq:positive_semideifinte_Q} as $U_{im} = (s_0\delta_{i,x} + s_1 \delta_{i,y} + s_2 \ii \delta_{i,x} + s_3 \ii  \delta_{i,y})v_m$ for $s_0, s_1,s_2,s_3\in \dsR$ and arbitrary $v_m$.
In this case, we have
\eqa{
& \sum_{mn} v_m^* \left[ s_0^2 Q_{x x}(\bsl{k}) + s_0 s_1 Q_{x y}(\bsl{k})  + s_0 s_2 \ii  Q_{x x}(\bsl{k}) + s_0 s_3 \ii  Q_{x y}(\bsl{k}) + s_1 s_0 Q_{y x}(\bsl{k}) + s_1^2 Q_{y y}(\bsl{k})  + s_1 s_2 \ii  Q_{y x}(\bsl{k}) + s_1 s_3 \ii  Q_{y y}(\bsl{k}) \right. \\
& \left. - \ii s_2 s_0 Q_{x x}(\bsl{k}) -\ii s_2  s_1 Q_{x y}(\bsl{k})  + s_2^2 Q_{x x}(\bsl{k}) +  s_2 s_3 Q_{x y}(\bsl{k}) - s_3 \ii s_0 Q_{y x}(\bsl{k}) - s_3 \ii  s_1 Q_{y y}(\bsl{k})  + s_2 s_3 Q_{y x}(\bsl{k}) +  s_3^2  Q_{y y}(\bsl{k})  \right]_{mn} v_n  \geq 0 \\
& \Rightarrow v^\dagger \left[ (s_0^2 +s_2^2) Q_{x x}(\bsl{k})  +  (s_1^2  + s_3^2) Q_{y y}(\bsl{k}) + (s_0 s_1 + s_2 s_3)(Q_{x y}(\bsl{k}) + Q_{y x}(\bsl{k})) +  (s_0 s_3 -s_1 s_2) \ii  (Q_{xy}(\bsl{k}) - Q_{yx}(\bsl{k}))\right] v  \geq 0  \\
& \Rightarrow v^\dagger \left[  (s_0^2 +s_2^2) G_{x x}(\bsl{k})  +    (s_1^2  + s_3^2)  G_{y y}(\bsl{k}) + ( s_0 s_1 + s_2 s_3)(G_{x y}(\bsl{k}) + G_{y x}(\bsl{k})) +(s_0 s_3 -s_1 s_2) F_{\bsl{k}}\right] v  \geq 0
}
for any complex vector $v$ and any $s_0, s_1,s_2,s_3\in \dsR$, where the last step uses \cref{eq:F_def,eq:G_def,eq:F_xy_in_short_as_F}.
Flipping the sign of $s_0$ and $s_1$ simultaneously, we obtain
\eq{
v^\dagger \left[  (s_0^2 +s_2^2) G_{x x}(\bsl{k})  +    (s_1^2  + s_3^2)  G_{y y}(\bsl{k}) + ( s_0 s_1 + s_2 s_3 )(G_{x y}(\bsl{k}) + G_{y x}(\bsl{k})) -  (s_0 s_3 -s_1 s_2)F_{\bsl{k}}\right] v  \geq 0 \ ,
}
which means
\eq{
\sum_l v^\dagger_l \left[  (s_0^2 +s_2^2) G_{x x}(\bsl{k})  +    (s_1^2  + s_3^2)  G_{y y}(\bsl{k}) + ( s_0 s_1 + s_2 s_3 )(G_{x y}(\bsl{k}) + G_{y x}(\bsl{k}))\right] v_l   \geq \sum_l \left| (s_0 s_3 -s_1 s_2) v^\dagger_l F_{\bsl{k}}v_l \right| 
}
for any complete orthonormal $v_l$ and any $s_0, s_1,s_2,s_3\in \dsR$.
We again choose $v_l$ to be eigenvectors of $F_{\bsl{k}}$, which leads to
\eq{
(s_0^2 +s_2^2)\left[g_{\bsl{k}}\right]_{x x}  +    (s_1^2  + s_3^2) \left[g_{\bsl{k}}\right]_{y y} + ( s_0 s_1 + s_2 s_3) (\left[g_{\bsl{k}}\right]_{x y} +\left[g_{\bsl{k}}\right]_{y x}) \geq \left| s_0 s_3 -s_1 s_2 \right|\rho(F_{\bsl{k}})
}
The above inequality infers that $g_{\bsl{k}} + \rho(F_{\bsl{k}}) \sigma_y/2$ is positive semi-definitive (with $\sigma_y$ here referring to the Pauli $y$ matrix), because
\eqa{
& \mat{ s_0 -\ii s_2 & s_1 -\ii s_4} \mat{ \left[g_{\bsl{k}}\right]_{x x}  & \left[g_{\bsl{k}}\right]_{x y}  - \ii \rho(F_{\bsl{k}})/2\\ \left[g_{\bsl{k}}\right]_{y x} + \ii \rho(F_{\bsl{k}})/2 &  \left[g_{\bsl{k}}\right]_{y y} } \mat{ s_0 +\ii s_2 \\ s_1 + \ii s_4} \\
& =(s_0^2 +s_2^2)\left[g_{\bsl{k}}\right]_{x x}  +    (s_1^2  + s_3^2) \left[g_{\bsl{k}}\right]_{y y} + ( s_0 s_1 + s_2 s_3) (\left[g_{\bsl{k}}\right]_{x y} +\left[g_{\bsl{k}}\right]_{y x}) + ( s_0 s_3 -s_1 s_2  )\rho(F_{\bsl{k}})  \geq 0 
}
holds for any $s_0, s_1,s_2,s_3\in \dsR$.
Then, we can derive the bound for $\det(g_{\bsl{k}})$:
\eqa{
& \det \mat{ \left[g_{\bsl{k}}\right]_{x x}  & \left[g_{\bsl{k}}\right]_{x y}  - \ii \rho(F_{\bsl{k}})/2\\ \left[g_{\bsl{k}}\right]_{y x} + \ii \rho(F_{\bsl{k}})/2 &  \left[g_{\bsl{k}}\right]_{y y} }  \geq 0  \Rightarrow  \det(g_{\bsl{k}}) - \rho(F_{\bsl{k}})^2/4  \geq 0  \Rightarrow  2\sqrt{\det(g_{\bsl{k}})} \geq \rho(F_{\bsl{k}}) \\
}
This bound is tighter than the lower bound of $\Tr[g_{\bsl{k}}]$ because
\eq{
\label{eq:g_lower_bound_by_rho_F}
\Tr[g_{\bsl{k}}] \geq 2\sqrt{\det(g_{\bsl{k}})} \geq \rho(F_{\bsl{k}})\ .
}

\section{Mathematical Concepts}
In this appendix, we discuss two mathematical concepts: non-abelian Stokes theorem and Schatten Norms.

\subsection{Non-abelian Stokes Theorem}
\label{app:non-abelian_stokes}

In this part, we review the non-abelian Stokes theorem following \cite{Arefeva1980nonabelianstokes}.
Before considering the generic case, let us first discuss the case of a rectangular region $D$ parametrized by $(x,y)$ with $x\in[0,X]$ and $y\in[0,Y]$.
Suppose we have $N$ vector fields (just like the periodic parts of $N$ Bloch states) defined on $D$, labeled as $v_{n,x,y}$ with $n=1,2,3,...,N$.
Here we require $v_{n,x,y}^\dagger v_{n',x,y} = \delta_{nn'}$, and we use $x,y$ instead of $\bsl{k}$ to show that the derivation holds for any paramtertization $x,y$, not just the Bloch momentum.
The corresponding projector reads
\eq{
P_{x,y} = v_{x,y} v^\dagger_{x,y}\ ,
}
where
\eq{
v_{x,y} = ( v_{1,x,y} \   v_{2,x,y}  \  v_{3,x,y} \  \cdots \  v_{N,x,y} )\ .
}
Clearly, we can define the Wilson line for $v$, which reads
\eq{
W(\gamma) = \lim_{L\rightarrow \infty}v_{x_0,y_0}^\dagger P_{x_1,y_1} P_{x_2,y_2} P_{x_3,y_3} \cdots P_{x_{L-1},y_{L-1}} v_{x_{L},y_{L}}\ ,
}
where $(x_0,y_0), (x_1,y_1), ..., (x_{L-1},y_{L-1}), (x_L,y_L)$ are arranged sequentially on the path $\gamma$, and $(x_0,y_0)$ and $ (x_L,y_L)$ are the initial and final points of $\gamma$, respectively.
We can also define the non-abelian Berry connection
\eq{
\label{eq:nonabelian_berry_connection}
\bsl{A}(x,y) = \ii ( v_{x,y}^\dagger  \partial_{x}  v_{x,y}, v_{x,y}^\dagger \partial_y v_{x,y} )
}
and the non-abelian Berry curvature
\eq{
\label{eq:nonabelian_berry_curvature}
F_{x,y} = \partial_x A_2(x,y) - \partial_y A_1(x,y) - \ii [ A_1(x,y) , A_2 (x,y) ]\ . 
}

To prove the relation between the Wilson loop and non-ablian Berry curvature, let us split the $[0,X]$ into $L_1$ equal parts and $[0,Y]$ into $L_2$ equal parts, resulting in area into $L_1\times L_2$ rectangular plaquettes for $D$.
A generic plaquette is labeled as 
\eq{
\widetilde{D}_{x_i,y_j}= \left\{ (x,y) | x \in [ x_i,x_i+dx] ,y \in [y_j , y_j +dy ] \right\} \ ,
}
where $x_i = i X /L_1$ with $i=0,1,2,...,L_1-1$, and $y_j = j Y /L_2$ with $j=0,1,2,...,L_2-1$, $dx = X/L_1$, and $dy = Y/L_2$.
Eventually, we will take $L_1$ and $L_2$ to infinity, which means $dx$ and $dy$ are infinitesimal quantities.

For the infinitesimal $\widetilde{D}_{x_i,y_j}$, the Wilson loop around the boundary of $\widetilde{D}_{x_i,y_j}$ is naturally connected to the non-abelian Berry curvature~\cite{fukui2005chern} via
\eqa{
\label{eq:w_ij}
w_{i,j} & = v_{x_i,y_j}^\dagger v_{x_i+dx,y_j}  v_{x_i+dx,y_j}^\dagger  v_{x_i+dx,y_j+dy} \left( v_{x_i,y_j+dy}^\dagger v_{x_i+dx,y_j+dy} \right)^{-1}  \left( v_{x_i,y_j}^\dagger v_{x_i,y_j+dy} \right)^{-1} \\
& = \left[ 1 -\ii A_1(x_i, y_j) dx + \frac{1}{2}v_{x_i,y_j}^\dagger \partial_{x_i}^2 v_{x_i,y_j}dx^2 \right] \\
& \quad \times \left[1-\ii A_2(x_i, y_j) dy - \ii\partial_{x_i} A_2(x_i, y_j) dx dy +   \frac{1}{2}v_{x_i,y_j}^\dagger \partial_{y_j}^2 v_{x_i,y_j}dy^2  \right] \\
& \quad \times \left[1 - \ii A_1(x_i, y_j) dx - \ii \partial_{y_j} A_1(x_i , y_j) dxdy + \frac{1}{2}   v_{x_i,y_j}^\dagger \partial_{x_i }^2 v_{x_i,y_j}  dx^2 \right]^{-1} \\
& \quad \times \left[1- \ii  A_2(x_i, y_j) dy + \frac{1}{2}  v_{x_i,y_j}^\dagger  \partial_{y_j}^2 v_{x_i,y_j}  dy^2 \right]^{-1} + O(dx^3, dx^2 dy, dx dy^2, dy^3)\\
& = 1 - \ii A_1(x_i, y_j) dx -\ii  A_2(x_i, y_j) dy +\ii A_1(x_i, y_j) dx +\ii A_2(x_i, y_j) dy   \\
& \quad  - A_1(x_i, y_j) A_2(x_i, y_j) dx  dy  + A_1(x_i, y_j)  A_2(x_i, y_j) dx  dy  \\ 
& \quad  + A_2(x_i, y_j) A_1(x_i, y_j) dxdy - A_1(x_i, y_j) A_2(x_i, y_j) dx dy \\
& \quad  -  \ii\partial_{x_i} A_2(x_i, y_j) dx dy +\ii \partial_{y_j} A_1(x_i , y_j) dxdy  +  O(dx^3, dx^2 dy, dx dy^2, dy^3) \\
& = 1  -  \ii\partial_{x_i} A_2(x_i, y_j) dx dy +  \ii \partial_{y_j} A_1(x_i , y_j) dxdy - [ A_1(x_i, y_j) , A_2(x_i, y_j)] dxdy  \\
& \quad + O(dx^3, dx^2 dy, dx dy^2, dy^3) \\
& = 1  -  \ii F_{x_i,y_j} dx dy   + O(dx^3, dx^2 dy, dx dy^2, dy^3)  \\
& = \exp\left[  -  \ii F_{x_i,y_j} dx dy  + O(dx^3, dx^2 dy, dx dy^2, dy^3) \right] \  ,
}
where we note that we have used the matrix inverse rather than the Hermitian conjugate in the first line.

To recover the non-abelian Berry curvature on the entire $D$ from the infinitesimal result in \cref{eq:w_ij}, one naïve way is to (matrix) multiply all $w_{ij}$ together. But as discussed in the Main Text, the non-abelian nature is incompatible with this guess. The solution is to dress $w_{ij}$:
\eq{
\widetilde{w}_{ij} = h_{i,j} w_{i,j} h_{i,j}^{-1} \ ,
}
where
\eq{
\label{eq:apphij}
h_{i,j} =   v_{x_0,y_0} v_{x_0,y_0}^\dagger v_{x_1,y_0} v_{x_1,y_0}^\dagger v_{x_2,y_0} \cdots   v_{x_{i-1},y_{0}}^\dagger v_{x_{i},y_0}  v_{x_i,y_0}^\dagger  v_{x_i,y_1} v_{x_i,y_2}^\dagger v_{x_i,y_3} \cdots v_{x_{i},y_{j-1}}^\dagger v_{x_{i},y_{j}}  \ ,
}
which has the following properties:
\eqa{
h_{i,j+1}  =  h_{i,j}  v_{x_{i},y_{j}}^\dagger v_{x_{i},y_{j+1}} 
}
and
\eq{
h_{i+1,0} = h_{i,0} v_{x_{i},y_0}^\dagger v_{x_{i+1},y_{0}}   \ .
}

Now we would like to derive the multiplication of all $\widetilde{w}_{i,j}$'s step by step.
We first consider the multiplication of $\widetilde{w}_{i,j}$ and $\widetilde{w}_{i,j+1}$ :
\eqa{
&  \widetilde{w}_{i,j} \widetilde{w}_{i,j+1}\\
& = h_{i,j}  v_{x_i,y_j}^\dagger v_{x_{i+1},y_j}  v_{x_{i+1},y_j}^\dagger  v_{x_{i+1},y_{j+1}} \left( v_{x_i,y_{j+1}}^\dagger v_{x_{i+1},y_{j+1}} \right)^{-1}  \left( v_{x_i,y_j}^\dagger v_{x_i,y_{j+1}} \right)^{-1}  h_{i,j}^{-1} \\
& \qquad \times h_{i,j}   v_{x_{i},y_{j}}^\dagger v_{x_{i},y_{j+1}} v_{x_{i},y_{j+1}}^\dagger v_{x_{i+1},y_{j+1}}  v_{x_{i+1},y_{j+1}}^\dagger  v_{x_{i+1},y_{j+2}} \left( v_{x_{i},y_{j+2}}^\dagger v_{x_{i+1},y_{j+2}} \right)^{-1}  \left( v_{x_{i},y_{j+1}}^\dagger v_{x_{i},y_{j+2}} \right)^{-1} \left[v_{x_{i},y_{j}}^\dagger v_{x_{i},y_{j+1}} \right]^{-1} h_{i,j}^{-1} \\
& = h_{i,j}  v_{x_i,y_j}^\dagger v_{x_{i+1},y_j}  \left( v_{x_{i+1},y_j}^\dagger  v_{x_{i+1},y_{j+1}}   v_{x_{i+1},y_{j+1}}^\dagger  v_{x_{i+1},y_{j+2}}\right) \\
& \qquad \times    \left( v_{x_{i},y_{j+2}}^\dagger v_{x_{i+1},y_{j+2}} \right)^{-1}  \left( v_{x_{i},y_{j}}^\dagger v_{x_{i},y_{j+1}} v_{x_{i},y_{j+1}}^\dagger v_{x_{i},y_{j+2}} \right)^{-1} h_{i,j}^{-1}  \ ;
}
then the multiplication of all $\widetilde{w}_{i,j}$'s with fixed $i$ reads
\eqa{
 \widetilde{u}_i & =\widetilde{w}_{i,0} \widetilde{w}_{i,1} \cdots    \widetilde{w}_{i,L_1-1} \\
& = h_{i,0} \left[v_{x_i,y_0}^\dagger v_{x_{i+1},y_0} \right]  \left[  v_{x_{i+1},y_{0}} ^\dagger   v_{x_{i+1},y_{1}} \cdots v_{x_{i+1},y_{L-2}} ^\dagger   v_{x_{i+1},y_{L-1}} v_{x_{i+1},y_{L-1}} ^\dagger   v_{x_{i+1},y_{L}}  \right]    \\
& \qquad \times \left[ v_{x_{i},y_L}^\dagger v_{x_{i+1},y_{L}} \right]^{-1} \left[  v_{x_{i},y_{0}} ^\dagger   v_{x_{i},y_{1}} \cdots v_{x_{i},y_{L-2}} ^\dagger   v_{x_{i},y_{L-1}} v_{x_{i},y_{L-1}} ^\dagger   v_{x_{i},y_{L}}  \right]^{-1}  h_{i,0}^{-1} \ .
}

Now we multiply $ \widetilde{u}_{i}$ by $ \widetilde{u}_{i+1}$, leading to
\eqa{
&\widetilde{u}_{i+1}  \widetilde{u}_{i}  \\
&  = h_{i,0}  v_{x_{i},y_0}^\dagger v_{x_{i+1},y_{0}} \left[v_{x_{i+1},y_0}^\dagger v_{x_{i+2},y_0} \right]  \left[  v_{x_{i+2},y_{0}} ^\dagger   v_{x_{i+2},y_{1}} \cdots v_{x_{i+2},y_{L-2}} ^\dagger   v_{x_{i+2},y_{L-1}} v_{x_{i+2},y_{L-1}} ^\dagger   v_{x_{i+2},y_{L}}  \right]    \\
& \qquad \times \left[ v_{x_{i+1},y_L}^\dagger v_{x_{i+2},y_{L}} \right]^{-1} \left[  v_{x_{i+1},y_{0}} ^\dagger   v_{x_{i+1},y_{1}} \cdots v_{x_{i+1},y_{L-2}} ^\dagger   v_{x_{i+1},y_{L-1}} v_{x_{i+1},y_{L-1}} ^\dagger   v_{x_{i+1},y_{L}}  \right]^{-1} \left[  v_{x_{i},y_0}^\dagger v_{x_{i+1},y_{0}} \right]^{-1} h_{i,0}^{-1}\\
& \qquad \times  h_{i,0} \left[v_{x_i,y_0}^\dagger v_{x_{i+1},y_0} \right]  \left[  v_{x_{i+1},y_{0}} ^\dagger   v_{x_{i+1},y_{1}} \cdots v_{x_{i+1},y_{L-2}} ^\dagger   v_{x_{i+1},y_{L-1}} v_{x_{i+1},y_{L-1}} ^\dagger   v_{x_{i+1},y_{L}}  \right]    \\
& \qquad \times \left[ v_{x_{i},y_L}^\dagger v_{x_{i+1},y_{L}} \right]^{-1} \left[  v_{x_{i},y_{0}} ^\dagger   v_{x_{i},y_{1}} \cdots v_{x_{i},y_{L-2}} ^\dagger   v_{x_{i},y_{L-1}} v_{x_{i},y_{L-1}} ^\dagger   v_{x_{i},y_{L}}  \right]^{-1}  h_{i,0}^{-1}\\
&  = h_{i,0}  \left[v_{x_{i},y_0}^\dagger v_{x_{i+1},y_{0}} v_{x_{i+1},y_0}^\dagger v_{x_{i+2},y_0} \right]  \left[  v_{x_{i+2},y_{0}} ^\dagger   v_{x_{i+2},y_{1}} \cdots v_{x_{i+2},y_{L-2}} ^\dagger   v_{x_{i+2},y_{L-1}} v_{x_{i+2},y_{L-1}} ^\dagger   v_{x_{i+2},y_{L}}  \right]    \\
& \qquad \times \left[v_{x_{i},y_L}^\dagger v_{x_{i+1},y_{L}} v_{x_{i+1},y_L}^\dagger v_{x_{i+2},y_{L}} \right]^{-1}  \left[  v_{x_{i},y_{0}} ^\dagger   v_{x_{i},y_{1}} \cdots v_{x_{i},y_{L-2}} ^\dagger   v_{x_{i},y_{L-1}} v_{x_{i},y_{L-1}} ^\dagger   v_{x_{i},y_{L}}  \right]^{-1}  h_{i,0}^{-1}\ .
}
Then, we can define the path-ordered multiplication of all $\widetilde{w}_{i,j}$'s:
\eqa{
\mathcal{P} \prod_{i,j} \widetilde{w}_{i,j}  = \prod_{i=L_1-1,...,2,1,0} \prod_{j=0,1,2,...,L_2-1}   \widetilde{w}_{i,j} \ ,
}
which in the limit of $L_1,L_2\rightarrow\infty$ reads
\eqa{
&  \lim_{L_1,L_2\rightarrow\infty}  \mathcal{P} \prod_{i,j} \widetilde{w}_{i,j}  \\
& =  \widetilde{u}_{L_1-1}\widetilde{u}_{L_1-2}\cdots \widetilde{u}_{0}  \\
& = \lim_{L_1,L_2\rightarrow\infty}   h_{0,0} \left[  v_{x_{0},y_{0}} ^\dagger   v_{x_{1},y_{0}} \cdots v_{x_{L_1-2},y_{0}} ^\dagger    v_{x_{L_1-1},y_{0}} v_{x_{L_1-1},y_{0}}^\dagger    v_{x_{L_1},y_{0}} \right]  \\
& \qquad \times \left[ v_{x_{L_1},y_0}^\dagger v_{x_{L_1},y_{1}} v_{x_{L_1},y_{1}}^\dagger v_{x_{L_1},y_{2}} \cdots v_{x_{L_1},y_{L_2-1}} ^\dagger    v_{x_{L_1},y_{L_2}}\right] \\
& \qquad \times \left[  v_{x_{0},y_{L_2}} ^\dagger   v_{x_{1},y_{L_2}} \cdots v_{x_{L_1-2},y_{L_2}} ^\dagger    v_{x_{L_1-1},y_{L_2}} v_{x_{L_1-1},y_{L_2}}^\dagger    v_{x_{L_1},y_{L_2}} \right]^{-1} \\
& \qquad \times \left[ v_{x_{0},y_0}^\dagger v_{x_{0},y_{1}} v_{x_{0},y_{1}}^\dagger v_{x_{0},y_{2}} \cdots v_{x_{0},y_{L_2-1}} ^\dagger    v_{x_{0},y_{L_2}}\right]^{-1} h_{0,0}^{-1}\\
& = W_{\partial D}\ ,
}
where we used $h_{0,0}=1$.

On the other hand, \cref{eq:w_ij} gives us 
\eqa{
\label{eq:w_tilde_ij}
& \widetilde{w}_{i,j} \\
& = h_{ij} \exp\left[  -  \ii F_{x_i,y_j} dx dy  + O(dx^3, dx^2 dy, dx dy^2, dy^3) \right]  h_{ij}^{-1}\\
& =  \exp\left[  -  \ii h_{ij} F_{x_i,y_j} h_{ij}^{-1} dx dy  + O(dx^3, dx^2 dy, dx dy^2, dy^3) \right]  \ ,
}
which means
\eqa{
\lim_{L_1,L_2\rightarrow\infty} \mathcal{P} \prod_{i,j} \widetilde{w}_{i,j} & = \lim_{L_1,L_2\rightarrow\infty} \prod_{i=L_1-1,...,2,1,0}  \prod_{j=0,1,2,...,L_2-1}  \widetilde{w}_{i,j} \\
& =  \lim_{L_1,L_2\rightarrow\infty} \prod_{i=L_1-1,...,2,1,0} \prod_{j=0,1,2,...,L_2-1}  \exp\left[  -  \ii h_{ij} F_{x_i,y_j} h_{ij}^{-1} dx dy  + O(dx^3, dx^2 dy, dx dy^2, dy^3) \right] \\
& \equiv \mathcal{P} \exp\left[ -  \ii \int_D dxdy \widetilde{F}_{x,y} \right] \ ,
}
where
\eq{
\label{eq:dressed_F}
\widetilde{F}_{x,y} = h_{x,y} F_{x,y} h_{x,y}^{-1}\ ,
}
and
\eq{
h_{x,y} = W((x_0,y_0)\rightarrow (x,y_0)) W((x,y_0)\rightarrow (x,y))
}
with $(x,y)\rightarrow (x',y')$ being the straight path from $(x,y)$ to $(x',y')$. Note that we use the indices to avoid clashing notation with \cref{eq:apphij}. 
Eventually, we arrive at the non-abelian stokes theorem
\eq{
\label{eq:non-abelian_stokes}
W(\partial D) = \mathcal{P} \exp\left[ - \ii \int_D \widetilde{F}_{x,y} dx dy\right]\ .
}

\cref{eq:non-abelian_stokes} holds for any simply connected $D$, since it is homeomorphic to a unit disk, which can be parameterized by two parameters with fixed range.

\subsection{Schatten Norms}
\label{app:schatten_norm}

In this part, we briefly review the definition of Schatten norms, which will be used in the proof of bounds, and provide some useful properties for them.

\subsubsection{General Definition}

Given a generic complex matrix $A$, its Schatten $p$-norm is defined as~\cite{bhatia2009positive} 
\eq{
\norm{A}_p = \left[  \sum_{i} s_i(A)^p \right]^{1/p}
}
with $p\in[1,\infty]$, where $s_i(A)$ is the $i$th singular value of $A$, and the summation is over all singular values of $A$. Recall that the singular values of a square matrix $A$ are the square roots of the eigenvalues of $AA^\dag$.
The Schatten $p$-norm is a matrix norm~\cite{bhatia2009positive}, which means it must have the following properties.
\begin{itemize}
\label{bulletprop}
    \item $\norm{A}_p \geq 0$;
    \item $\norm{A}_p = 0 \Leftrightarrow A = 0 $;
    \item $ \norm{\alpha A}_p = |\alpha | \norm{A}_p$ for any $\alpha \in \dsC$;
    \item $\norm{A+B}_p\leq \norm{A}_p + \norm{B}_p$ for any matrix $B$ that has the same dimension as $A$.
\end{itemize}

\subsubsection{Schatten $1$-norm for Hermitian Matrices}

In particular, we will use the Schatten $1$-norm for hermitian matrix $A$.
For convenience, we define
\eq{
\rho(A) = \norm{A}_1\ .
}
For Hermitian matrices, singular values are the same as the absolute values of the eigenvalues, and thus the summation of the singular values in the norm can be replaced by the summation of the absolute values of the eigenvalues, \ie,
\eq{
\rho(A) = \norm{A}_1 = \sum_{i} s_i(A)  = \sum_{i} |\lambda_i| \ ,
}
where $\lambda_i$ is the $i$th eigenvalue of the Hermitian $A$.

There are two useful properties of the Schatten $1$-norm $\rho$ for Hermitian matrix $A$.
First, given a  generic set of othonormal set of vectors $u_{\alpha}$, then we have
\eq{
\rho(A) \geq \sum_{\alpha} \left| u_{\alpha}^\dagger A u_{\alpha} \right|\ .
}
To see that, suppose $\{ v_i \}$ is the complete set of orthonormal eigenvectors of $A$ for eigenvalue $\lambda_i$, and we have
\eq{
C_{i\alpha} = v_i^\dagger u_\alpha
}
with $\sum_{\alpha} |C_{i,\alpha}|^2\leq 1$ where the equality holds when $u_\alpha$ is complete.
Then, 
\eq{
\sum_{\alpha}\left| u_{\alpha}^\dagger A u_{\alpha} \right| = \sum_{\alpha}\left| \sum_{i} C_{i\alpha}^* C_{i\alpha} \lambda_i \right| = \sum_{\alpha}\left| \sum_{i} |C_{i\alpha}|^2 \lambda_i \right| \leq \sum_{\alpha}\sum_{i} |C_{i\alpha}|^2 \left| \lambda_i \right| \leq \sum_{i}  \left| \lambda_i \right| = \rho(A)\ .
}

Second, consider a hermitian matrix that smoothly depends on $s$, \ie, $A(s)$.
Suppose $\{v_{i,s} \}$ is the complete set of orthonormal eigenvectors of of $A(s)$ with eigenvalues labeled by $\lambda_{i,s}$, we have
\eq{
\label{eq:linearineq}
\rho\left(\frac{dA(s)}{ds}\right) \geq \sum_{i} \left|v^\dagger_{i,s} \frac{dA(s)}{ds} v_{i,s} \right| = \sum_{i}  \left|\frac{d (v^\dagger_{i,s} A(s) v_{i,s} ) }{ds}\right| =  \sum_{i}  \left|\frac{d \lambda_{i,s} }{ds}\right| \ ,
}
where we consider the case that $\lambda_{i,s}$ is isolated in a neighborhood of $s$. Here we used the Feynman-Hellman theorem  $v^\dagger_{i,s} \frac{dA(s)}{ds} v_{i,s} = \frac{d}{ds}(v^\dagger_{i,s} A(s) v_{i,s})$ which is locally valid since we can always choose a smooth gauge for $v_i$ in the neighborhood of $s$. Note that the final inequality is completely gauge-invariant.
To make contact with the Wilson loop, we will need an exponentiated form of the inequality \ref{eq:linearineq}:
\eq{
\label{eq:rho_derivative_matrix}
\rho\left(-\ii   \frac{d}{ds} \left[ e^{
    \ii A(s)}  \right] e^{-\ii A(s)} \right)  \geq \sum_{i}\left| -\ii   v^\dagger_{i,s}  \frac{d}{ds} \left[ e^{
    \ii A(s)}  \right] e^{-\ii A(s)} v_{i,s} \right| = \sum_{i}\left| -\ii   v^\dagger_{i,s}  \frac{d}{ds} \left[ e^{
    \ii A(s)}  \right] v_{i,s} e^{-\ii \lambda_{i,s}}  \right| = \sum_{i}\left|  \frac{d\lambda_{i,s}}{ds}  \right|
}
Moreover, owing to 
\eq{
\frac{d}{ds} e^{X(s)} = \int_0^1 e^{\alpha X(s)} \frac{d X(s)}{ds} e^{(1-\alpha) X(s)} d\alpha\ ,
}
we have
\eq{
\rho\left(-\ii   \frac{d}{ds} \left[ e^{
    \ii A(s)}  \right] e^{-\ii A(s)} \right) = \rho\left (  \int_0^1 e^{\ii \alpha A(s)} \frac{d A(s)}{ds} e^{-\alpha \ii A(s)} d\alpha \right) \leq   \int_0^1 \rho\left (e^{\ii \alpha A(s)} \frac{d A(s)}{ds} e^{-\alpha \ii A(s)}\right) d\alpha = \rho(\frac{d A(s)}{ds})
}
where we have used the triangle inequality of the Schatten norm. 
In sum, we have
\eq{
\label{eq:rho_derivative_unitary}
\sum_{i}\left|\frac{d \lambda_{i,s} }{ds}\right| \leq \rho\left(-\ii   \frac{d}{ds} \left[ e^{
    \ii A(s)}  \right] e^{-\ii A(s)} \right) \leq \rho(\frac{d A(s)}{ds})\ , 
}
for the eigenvalues $\lambda_{i,s}$ of $A(s)$, which means the exponentiated form is a tighter bound than \cref{eq:linearineq}.

Immediately we have, given two Hermitian matrices $A(s)$ and $B(s)$ that smoothly depend on $s$, we have
\eqa{
    \label{eq:norm_derivative_lemma}
  & \rho\left(-\ii   \frac{d}{ds} \left[ e^{
    \ii A(s)} e^{\ii B(s)} \right]  e^{-\ii B(s)} e^{-\ii A(s)} \right) = \rho\left(-\ii \left[\frac{d}{ds}  e^{
    \ii A(s)}\right] e^{-\ii A(s)} -\ii  e^{
    \ii A(s)}  \left[\frac{d}{ds}   e^{\ii B(s)} \right] e^{-\ii B(s)}  e^{-\ii A(s)} \right) \\
 & \leq  \rho\left(-\ii  \left[\frac{d}{ds}  e^{
    \ii A(s)}\right] e^{-\ii A(s)}\right) +\rho\left( -\ii   e^{
    \ii A(s)}  \left[\frac{d}{ds}   e^{\ii B(s)} \right] e^{-\ii B(s)}  e^{-\ii A(s)} \right)  \\
    & = \rho\left(-\ii \left[\frac{d}{ds}  e^{
    \ii A(s)}\right] e^{-\ii A(s)}\right) +\rho\left( -\ii  \left[\frac{d}{ds}   e^{\ii B(s)} \right] e^{-\ii B(s)}  \right) \\
    & \leq \rho\left ( \frac{d A(s)}{ds} \right)  +  \rho\left ( \frac{d B(s)}{ds} \right)\ ,
}
where we, for the third line, use the fact that unitary transformations cannot change eigenvalues.

\section{Wilson loop Lower Bound of the Integrated Quantum Metric}
\label{app:WL_bound}

In this appendix, we prove the WL lower bound of the integrated quantum metric for an isolated set of $N$ bands in 2D.

\begin{figure}[t]
    \centering
    \includegraphics[width=0.8\linewidth]{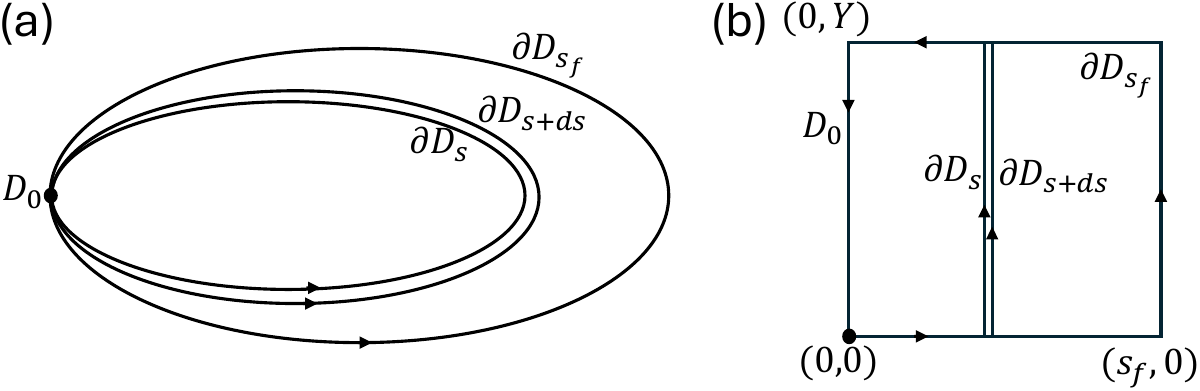}
    \caption{
    These is two examples of the $D_s$ with $D_0$.
    In (a), $D_0$ is the black point, and $\partial D_{s>0}$ are the black solid lines.
    In this case, $D_0$ is the starting point for $W(\partial D_s)$ for all $s$.
    In (b), $D_s=[0,s]\times [0,Y]$.
    The black dot in this case is not $D_0$ but still the starting point for $W(\partial D_s)$ for all $s$.
    }
    \label{fig:D_s}
\end{figure}

Consider a simply connected region that depend smoothly on a continuous parameter $s\in[0,s_f]$, labeled as $D_s$, with $D_0$ having zero area.
We require that $D_s\subset D_{s'}$ for any $s\leq s'$, and require the starting point $\bsl{k}_0$ of $W(\partial D_s)$ is the same for all $s\in[0,s_f]$.
We refer to this $D_s$ as a proper deformation.
(See two examples of proper $D_s$ in \cref{fig:D_s}.)
The eigendecomposition of $W(\partial D_s)$ reads $W(\partial D_s) = V_s e^{\ii \Lambda(s)} V_s^\dagger$ with real diagonal continuous $\Lambda(s) = \diag(...,\lambda_{i,s},...)$ and unitary $V_s$.
We choose $\Lambda(s=0) = 0$ without loss of generality, since $W(\partial D_0) = 1$ always holds.
Then, we have
\eqa{
\label{eq:ineq_absolute_winding}
& \int_0^{s_f} ds \  \rho\left( \left[ -\ii \partial_{s} W(\partial D_s) \right] W(\partial D_s)^\dagger \right) =  \int_0^{s_f} ds \  \rho\left( W(\partial D_s)^\dagger\left[ -\ii \partial_{s} W(\partial D_s) \right]  \right) =  \\
& \qquad \int_0^{s_f} ds \  \rho\left(\partial_{s}\left[ -\ii e^{\ii V_s  \Lambda(s) V_s^\dagger  }\right]   e^{-\ii V_s  \Lambda(s) V_s^\dagger  }  \right) \geq  \int_0^{s_f} ds \sum_{i} \left| \frac{d\lambda_{i,s}}{d s}\right|  \ ,
}
where we have used \cref{eq:rho_derivative_unitary}.
We call $\int_0^{s_f} ds \sum_{i} \left| \frac{d\lambda_{i,s}}{d s}\right|$ the absolute winding of the WL $W(\partial D_s)$.
We emphasize that the absolute winding does not depend on how we rank $\lambda_{i,s}$ as long as they are continuous.
We note that since $\lambda_{i,s}$ at most touches at measure zero points, we don't need to consider them when deriving \cref{eq:ineq_absolute_winding}, allowing us to use \cref{eq:rho_derivative_unitary} safely.

The next step is to show the absolute WL winding bounds the integrated quantum metric from below.
To show this, we first note that
\eqa{
&  \left[ -\ii \partial_{s} W(\partial D_s) \right]W(\partial D_s)^\dagger = -\ii  \frac{1}{ds} \left[ W(\partial D_{s+ds}) - W(\partial D_{s}) \right] W(\partial D_s)^\dagger  = -\ii  \frac{1}{ds} \left[ W(\partial D_{s+ds}) W(\partial D_s)^\dagger-1 \right] \ .
}
As shown in \cref{fig:D_s}, $W(\partial D_{s+ds})W(\partial D_s)^\dagger$ is a WL along the boundary of $D_{s+ds} -D_s $.
To simplify $W(\partial D_{s+ds})W(\partial D_s)^\dagger$, let us parametrize $\partial D_s$ as $\{ \bsl{k}(s,t) | t\in [0,1] \}$, where $\bsl{k}(s,0) = \bsl{k}(s,1) = \bsl{k}_0$ and $\bsl{k}(s,t)$ goes through $\partial D_s$ smoothly as $t$ increases from 0 to 1.
In this case, $D_{s+ds} -D_s  = \{ \bsl{k}(s',t) | t\in [0,1] , s'\in [ s, s+ds]\}$.
$ W(\partial D_{s+ds}) W(\partial D_s)^\dagger$ is then Wilson loop along the boundary of $(D_{s+ds} -D_s)$, as 
\eq{
 W(\partial D_{s+ds}) W(\partial D_s)^\dagger=W((s,0)\rightarrow (s+ds,0)) W((s+ds,0)\rightarrow (s+ds,1)) W((s+ds,1)\rightarrow (s,1))  W((s,1)\rightarrow (s,0)) \ ,
}
where we have used that $W((1,s+ds)\rightarrow (1,s))=W((0,s)\rightarrow (0,s+ds))=1$.
To use the non-abelian stokes theorem in \cref{eq:non-abelian_stokes}, we first note that the starting point of $W(\partial D_{s+ds})W(\partial D_s)^\dagger$ is $(s,0)$.
Define 
\eq{
h_{s',t} = W((s,0)\rightarrow (s',0)) W((s',0)\rightarrow (s',t))=  W((0,0)\rightarrow (s',0))  W((s',0)\rightarrow (s',t))\ ,
}
and 
\eq{
\widetilde{F}_{s',t} = h_{s',t} F_{s',t} h_{s',t}^{-1}\ ,
}
where $F_{s',t} $ is given by replacing $(x,y)$ in \cref{eq:nonabelian_berry_curvature} by $(s',t)$, and we have used
\eq{
 W((s,0)\rightarrow (s',0))  =  W((0,0)\rightarrow (s',0))=1\ .
}
Then, we have
\eqa{
& W(\partial D_{s+ds})W(\partial D_s)^\dagger \\
& = \mathcal{P}\exp\left[ -\ii \int_{s}^{s+ds} ds' \int_0^1 dt  \widetilde{F}_{s',t}\right] \\
& = \lim_{L_2\rightarrow \infty} \prod_{j=0}^{L_2-1} \exp\left[ -\ii ds \frac{1}{L_2}   \widetilde{F}_{s,j/L_2}\right]  + O(ds^2)\\
& = 1 - \ii ds  \int_0^1 dt\ \widetilde{F}_{(s,t)} + O(ds^2)\ ,
}
leading to
\eq{
\label{eq:W_dag_partial_W}
\left[ -\ii \partial_{s} W(\partial D_s) \right] W(\partial D_s)^\dagger  = -\int_0^1 dt\ \widetilde{F}_{(s,t)}\ .
}
Substituting \cref{eq:W_dag_partial_W} into the left-most term in \cref{eq:ineq_absolute_winding}, we obtain
\eqa{
\int_0^{s_f} ds  \ \rho\left( \left[ -\ii \partial_{s} W(\partial D_s) \right] W(\partial D_s)^\dagger \right) & = \int_0^{s_f} ds\  \rho\left( \int_0^1 dt\  \widetilde{F}_{s,t} \right)  \leq \int_0^{s_f} ds\int_0^1 dt\ \ \rho\left(  \widetilde{F}_{s,t} \right) = \int_0^{s_f} ds\int_0^1 dt\ \ \rho\left(  F_{s,t} \right) \  ,
}
where we use the fact that $F_{s,t}$ and $\widetilde{F}_{s,t}$ have the same eigenvalues.

Previously, we choose the parametrize $(s,t)$ to make sure the region of the parameters is rectangular, allowing us to directly use \cref{eq:non-abelian_stokes}.
We now convert it back to the Bloch momentum, which is what we normally use.
Define $F_{\bsl{k}}$ as \cref{eq:nonabelian_berry_curvature} with $(x,y)=(k_x,k_y)$.
Then,
\eq{
 F_{s,t} = \det\left[ \frac{\partial(k_x,k_y)}{\partial(s,t)} \right]  F_{\bsl{k}} \ ,
}
where $ \frac{\partial(k_x,k_y)}{\partial(s,t)} $ is the Jacobian matrix.
Then,
\eqa{
\int_0^{s_f} ds\int_0^1 dt\ \ \rho\left(  F_{s,t} \right) & = \int_0^{s_f} ds\int_0^1 dt\ \ \rho\left( \det\left[ \frac{\partial(k_x,k_y)}{\partial(s,t)} \right]   F_{\bsl{k}} \right) = \int_0^{s_f} ds\int_0^1 dt\ \ \left| \det\left[ \frac{\partial(k_x,k_y)}{\partial(s,t)} \right]  \right| \rho\left(   F_{\bsl{k}} \right)\\
& = \int_0^{s_f} ds\int_0^1 dt\ \ \left| \det\left[ \frac{\partial(k_x,k_y)}{\partial(s,t)} \right]  \right| \rho\left(   F_{\bsl{k}} \right) = \int_{D_{s_f}} d^2 k\ \rho\left(   F_{\bsl{k}} \right)\ ,
}
which means
\eq{
\int_0^{s_f} ds \  \rho\left( \left[ -\ii \partial_{s} W(\partial D_s) \right] W(\partial D_s)^\dagger \right)  \leq \int_{D_{s_f}} d^2 k \rho\left(  F_{\bsl{k}} \right) \ .
}

As \cref{eq:g_lower_bound_by_rho_F} suggests that $\Tr[g_{\bsl{k}}] \geq 2\sqrt{\det(g_{\bsl{k}})} \geq \rho(F_{\bsl{k}})$, we  arrive at 
\eq{
\label{eq:g_WL_bound}
\int_{D_{s_f}} d^2 k \Tr[g_{\bsl{k}}] \geq \int_{D_{s_f}} d^2 k\ 2\sqrt{\det(g_{\bsl{k}})}  \geq \int_0^{s_f} ds   \rho\left( \left[ -\ii \partial_{s} W(\partial D_s) \right] W(\partial D_s)^\dagger \right)\ \geq \int_0^{s_f} ds \sum_{i} \left| \frac{d\lambda_{i,s}}{d s}\right| \ .
}
$\int_0^{s_f} ds \sum_{i} \left| \frac{d\lambda_{i,s}}{d s}\right|$ is the absolute WL winding.

More importantly, there is an freedom in defining the absolute WL winding while keeping \cref{eq:g_WL_bound} valid.
Let us define $\mathcal{W}_{s}$ such that
\eq{
\label{eq:W_s}
W(\partial D_s) =  U_s \mathcal{W}_{s} U_s^\dagger   V
}
with unitary $V$ and $U_s$ and $U_s$ depending on $s$ smoothly, and define
$\phi_{l}(s)$ to be the phase of the $l$th eigenvalue of $\mathcal{W}_{s}$.
As long as we choose $\phi_{l}(s)$ to be continuous, we have
\eqa{
\label{eq:WL_lower_bound_of_Trg_for_Z_2}
\mathcal{N} & \equiv \int_0^{s_f} ds \sum_{l} \left|\frac{d\phi_{l}(s)}{ds} \right| \leq \int_{D_{s_f}}  d^2 k 2\sqrt{\det(g_{\bsl{k}})} \leq \int_{D_{s_f}}  d^2 k \Tr[g_{\bsl{k}}] \ ,
}
where $\mathcal{N}$ is called the absolute WL winding of the proper deformation $D_s$ and dressing $V$ and $U_s$.
\cref{eq:WL_lower_bound_of_Trg_for_Z_2} comes from the fact that $V$ is independent of $s$ and $ U_s \W_s U_s^\dagger$ have the same eigenvalues as $\W_s$, which gives
\eqa{
& \rho\left[ -\ii \partial_{s} \log W(\partial D_s) \right] = \rho\left( \left[ -\ii \partial_{s} W(\partial D_s) \right] W(\partial D_s)^\dagger \right)   = \rho\left( \left[ -\ii \partial_{s}(U_s \mathcal{W}_{s} U_s^\dagger   V) \right]  V^\dagger U_s \mathcal{W}_{s}^\dagger  U_s^\dagger   \right)  \\
&  = \rho\left( \left[ -\ii \partial_{s}(U_s \mathcal{W}_{s} U_s^\dagger   ) \right]  U_s \mathcal{W}_{s}^\dagger  U_s^\dagger   \right)   \geq \int_0^{s_f} ds \sum_{l} \left|\frac{d\phi_{l}(s)}{ds} \right|\ .
}
Combined with \cref{eq:g_WL_bound}, we arrive at \cref{eq:WL_lower_bound_of_Trg_for_Z_2}.

One special yet useful case for the WL winding bound is when $D_{s} = \{ k_1 \bsl{b}_1/(2\pi) + k_2 \bsl{b}_2/(2\pi) | k_1\in[0,s]\ , k_2\in[0,2\pi] \}$. (See \cref{fig:D_s}(b) with $Y=2\pi$.)
Here $\bsl{b}_1$ and $\bsl{b}_2$ are two basis reciprocal lattice vectors, which means $D_{2\pi}=\text{BZ}$ is the entire first Brillouin zone.
In this case, we can define a new WL which is 
\eq{
\label{eq:WL_Z2}
\mathcal{W}_{k_1} = W\left( k_1 \bsl{b}_1/ (2\pi) \rightarrow k_1 \bsl{b}_1/ (2\pi) + \bsl{b}_2 \right)\ ,
}
which is the WL along $k_2$ at a fixed $k_1$.
Then,  we have \cref{eq:WL_lower_bound_of_Trg_for_Z_2} for $\phi_{l}(s)$ being the continuous phase of the $l$th eigenvalue of $\mathcal{W}_{k_1 = s}$.
To prove this, we first note that 
\eqa{
W\left(\partial D_{s}\right)  = W\left(\Gamma \rightarrow s \bsl{b}_1/ (2\pi)\right)  \mathcal{W}_{s} W^{\dagger}\left(\Gamma \rightarrow s \bsl{b}_1/ (2\pi)\right) W\left(  \bsl{b}_2 \rightarrow \Gamma\right)\ ,
}
where $\Gamma = (0,0)$.
Compared to \cref{eq:W_s}, we clearly see that $W\left(\partial D_{s}\right)$ has the same expression in \cref{eq:W_s} with $U_s = W\left(\Gamma \rightarrow s \bsl{b}_1/ (2\pi)\right) $ and $V= W\left(  \bsl{b}_2 \rightarrow \Gamma\right)$.
Then, \cref{eq:WL_lower_bound_of_Trg_for_Z_2}  naturally follows.
%

\section{$\dsZ_2$ Lower Bound of the Integrated Quantum Metric}
\label{app:Z2_bound}

In this appendix, we derive the $\dsZ_2$ lower bound of the integrated quantum metric.
We will discuss two $\dsZ_2$ indices, one protected by the TR symmetry and the other protected by the particle-hole (PH) symmetry, which are in fact equivalent.

\subsection{Review of TR $\dsZ_2$ index from Wilson Loop}

Before discussing the lower bound, let us first review how to calculate TR $\dsZ_2$ index from WL.
Since the TR $\dsZ_2$ index is defined for 2D systems with spinful TR symmetry~\cite{Kane2005Z2}, we consider an isolated set of $2N$ bands owing to Kramer's degeneracy.
The WL of interest is $\mathcal{W}_{k_1}$ in \cref{eq:WL_Z2}.
Because of TR symmetry, we have
\eq{
\mathcal{W}_{k_1} = U_{\frac{k_1}{2\pi} \bsl{b}_1}^T \mathcal{W}_{k_1}^T U_{\frac{k_1}{2\pi} \bsl{b}_1}^*  \text{ for $k_1 = 0,\pi$\ ,} 
}
where $ \left[ U_{\bsl{k}} \right]_{mn} = \bra{ u_{-\bsl{k},m} }\TR \ket{u_{\bsl{k},n}  }$.
It means that $\mathcal{W}_0$ and $\mathcal{W}_{\pi}$ have Kramer's degeneracy.

As discussed in \refcite{Dai2011Z2WilsonLoop}, we track the phase of the eigenvalues of $\mathcal{W}_{k_1}$ from $k_1=0$ to $k_1=\pi$ to determine $\dsZ_2$ index.
Specifically, $\mathcal{W}_{k_1}$ has eigenvalues $e^{\ii \phi_l(k_1)}$ with $l=1,2,...,2N$, and we can choose $\phi_l(k_1)$ to be continuous for $k_1\in [0,\pi]$ which is always allowed.
Without loss of generality, we fix $\phi_l(0)\in [0,2\pi)$; furthermore, we have in general
\eq{
\phi_l(\pi) = \left[\phi_l(\pi)\ \mod\ 2\pi \right] + 2\pi M_l \ ,
}
where
\eq{
\left[ x\ \mod\ 2\pi \right]\in [0,2\pi)\ ,
}
and $M_l\in\dsZ$.
Finally, $\dsZ_2$ reads
\eq{
\label{eq:Z_2_WL_winding}
\dsZ_2 = \sum_{l=1}^{2N} M_l\ \mod\ 2 \ .
}

\subsection{Review of PH $\dsZ_2$ index}

As discussed in \refcite{Song2020TBGII}, the PH symmetry also protects a $\dsZ_2$ index in a same way as the TR symmetry, which we review in this section.

Consider the PH matrix $U_{\P}(\bsl{k}$), which satisfies $U_{\P}(\bsl{k}) h^*(\bsl{k}) U_{\P}^{\dagger}(\bsl{k}) = - h(-\bsl{k})$ and $U_{\P}(\bsl{k}) U_{\P}^*(-\bsl{k}) = -1$, where $h(\bsl{k})$ is the matrix Hamiltonian.
Suppose we have an isolated set of $2N$ bands of $h(\bsl{k})$ that preserves the PH symmetry, and we note the corresponding eigenvectors as $v_{n,\bsl{k}}$ with $n=1,2,3,...,2N$, which satisfies
\eq{
\label{eq:PH_band_basis}
U_{\P}(\bsl{k}) v_{n,\bsl{k}}^* = \sum_{m=1}^{2N} v_{m,-\bsl{k}} \left[U_{\bsl{k}}\right]_{mn} 
}
with unitary $U_{\bsl{k}}$.
Since \cref{eq:PH_band_basis} has the same form as the TR symmetry acting on the eigenstates, the PH-preserving $2N$ bands must be able to have the same $\dsZ_2$ topology as the TR-preserved bands, characterized by the same WL winding as \cref{eq:Z_2_WL_winding}.

\subsection{Proof of the $\dsZ_2$ Lower Bound of the Integrated Quantum Metric}

We will only show the derivation for the TR $\dsZ_2$, since the derivation for the PH $\dsZ_2$ is exactly the same.

According to \cref{eq:WL_lower_bound_of_Trg_for_Z_2}, we can find a $\dsZ_2$ lower bound of the integrated quantum metric as long as we can bound the absolute winding of $\mathcal{W}_{k_1}$.
In the following, we will prove
\eq{
\label{eq:winding_Z2_bound}
\sum_{l}\left| \int_0^\pi d k_1 \frac{d}{d k_1}  \phi_{l}(k_1) \right| \geq 2 \pi \dsZ_2 \ ,
}
where $\phi_{l}(k_1)$ are the phases of the eigenvalues of $\mathcal{W}_{k_1}$, which we choose to be continuous.
Of course, \cref{eq:winding_Z2_bound} is trivially true for $\dsZ_2 = 0$.
In the following, we will consider the case where $\dsZ_2=1$.

\begin{figure}
    \centering
    \includegraphics[width=0.9\linewidth]{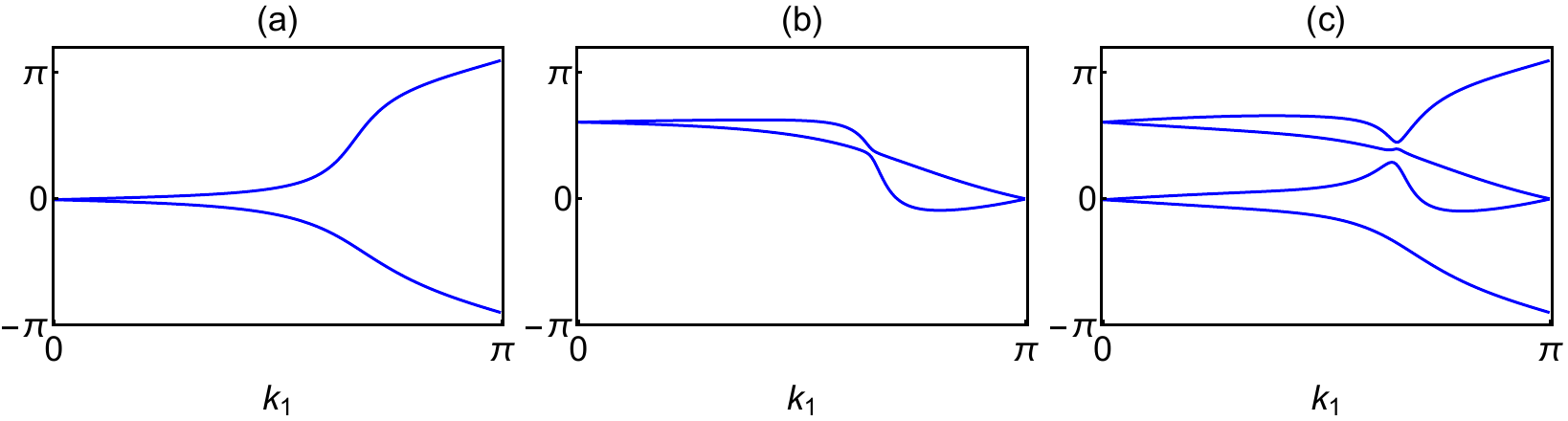}
    \caption{(a) A representative Wilson loop for an isolated set of two bands with $\dsZ_2 = 1$.
    (b) A representative Wilson loop for an isolated set of two bands with $\dsZ_2 = 0$.
    (c) A representative Wilson loop given by coupling (a) to (b), which is an isolated set of four bands with $\dsZ_2 = 1$.
    }
    \label{fig:Z2WLConnectivity}
\end{figure}

Before proving the general case, let us look at an example.
As shown by \refcite{Dai2011Z2WilsonLoop}, for an isolated set of two bands with $\dsZ_2 = 1$, the two WL bands are connected as shown in \cref{fig:Z2WLConnectivity}(a).
Since $\dsZ_2$ is a stable topology, it means that if we couple an isolated set of two bands with $\dsZ_2 = 1$ to an isolated set of two bands with $\dsZ_2 = 0$ (\cref{fig:Z2WLConnectivity}(b)), we obtain an isolated set of four bands with $\dsZ_2 = 1$.
As shown in \cref{fig:Z2WLConnectivity}(c), the isolated set of four bands with $\dsZ_2 = 1$ still have fully connected WL bands.
Therefore, this example suggests an intuitive argument that for an isolated set of any number of bands with $\dsZ_2 = 1$, the WL bands are fully connected.
Such argument was presented in \refcite{Song2020TBGII}.

Now we prove the connectivity for the WL bands in the case of $\dsZ_2 = 1$, which is essential for the proof of the $\dsZ_2$ bound.
Since the absolute WL winding does not care about how we rank $\phi_{l}(k_1)$, we choose 
\eq{
\label{eq:no_crossing_conditions}
\left[ \phi_{l}(k_1)-\phi_{l'}(k_1) + 2\pi n  \right] \left[ \phi_{l}(k_1')-\phi_{l'}(k_1') + 2\pi n  \right] \geq 0\ \text{for any $k_1,k_1'\in [0,\pi]$, $n\in\dsZ$ and different $l$ and $l'$}
}
and $\phi_l(0)\in [0,2\pi)$.
In other words, we make sure $\phi_{l}(k_1)$ and $\phi_{l'}(k_1)$ do not cross each other for any different $l$ and $l'$ and for any $2\pi$ shifts, though $\phi_{l}(k_1)$ and $\phi_{l'}(k_1)$ may still have touching that is either accidental or protected by extra symmetries for any $k_1\in (0,\pi)$.
\eqnref{eq:no_crossing_conditions} is always allowed.
Without loss of generality, we choose $\phi_{1}(k_1)\leq \phi_{2}(k_1)\leq \phi_{3}(k_1) \leq ... \leq \phi_{2N}(k_1)$.
What we want to show is that it is impossible to have a direct gap among $\phi_{l}(k_1)$'s, \ie, it is impossible to have $l_0$ that satisfies $\phi_{l_0+1}(k_1) > \phi_{l_0}(k_1)$ for all $k_1 \in [0,\pi]$.
First note that $e^{\ii \phi_l (\pi)}$ form Kramers pairs owing to TR symmetry.
In principle, given a Kramers pair $e^{\ii \phi_{l_1} (\pi)}$ and $e^{\ii \phi_{l_2} (\pi)}$ with $l_1\neq l_2$, only $\phi_{l_1} (\pi)=\phi_{l_2} (\pi)\ \mod\ 2\pi$ is required, whereas it is allowed to have $\phi_{l_1} (\pi)\neq \phi_{l_2} (\pi)$.
It is impossible to have $\phi_{l_1} (\pi) =\phi_{l_2} (\pi) $ for all Kramers pair  $e^{\ii \phi_{l_1} (\pi)}$ and $e^{\ii \phi_{l_2} (\pi)}$ with $l_1\neq l_2$, since otherwise $M_{l_1} = M_{l_2}$ in \cref{eq:Z_2_WL_winding} for all pairs and $\dsZ_2$ must be zero.
Therefore, we must have at least one Kramers pair $e^{\ii \phi_{l_1} (\pi)}$ and $e^{\ii \phi_{l_2} (\pi)}$ with $l_1\neq l_2$ such that $\phi_{l_1} (\pi) \neq \phi_{l_2} (\pi) $, which we call nontrivial Kramers pair.
Consider a generic nontrivial Kramers pair $e^{\ii \phi_{l_1} (\pi)}$ and $e^{\ii \phi_{l_2} (\pi)}$ with $l_1\neq l_2$.
We can assume $l_1<l_2 $ without loss of generality, which means  $\phi_{l_2} (\pi) =   \phi_{l_1} (\pi) + 2\pi n$ with integer $n>0$.
If $\phi_{l_2} (\pi)  < \phi_{2N} (\pi) $, then $\phi_{l_1} (k_1) + 2\pi n$ crosses $\phi_{2N} (k_1) $ since $\phi_{l_1} (0)+ 2\pi n > \phi_{2N} (0)$ and $\phi_{l_1} (\pi)+ 2\pi n < \phi_{2N} (\pi)$, which contradicts \cref{eq:no_crossing_conditions}.
Similarly, if $\phi_{l_1} (\pi)  > \phi_{1} (\pi) $, then $\phi_{l_2} (k_1) - 2\pi n$ crosses $\phi_{1} (k_1) $ since $\phi_{l_2} (0)- 2\pi n < \phi_{2N} (0)$ and $\phi_{l_2} (\pi)- 2\pi n > \phi_{0} (\pi)$, which again contradicts \cref{eq:no_crossing_conditions}.
Therefore, for any nontrivial Kramers pair $e^{\ii \phi_{l_1} (\pi)}$ and $e^{\ii \phi_{l_2} (\pi)}$ with $l_1< l_2$, we have $\phi_{2N} (\pi) =  \phi_{l_2} (\pi) = \phi_{l_1} (\pi) + 2\pi n  = \phi_{0} (\pi) + 2 \pi n $ with integer $n>0$, meaning that $M_{l_1}+M_{l_2} = 2 M_{l_1} + n$ and $n$ is the same for different nontrivial Kramers pair.

In other words, suppose we have $x$ nontrivial Kramers pair, the $\dsZ_2$ index is just $x n\ \mod\ 2$, according to \cref{eq:Z_2_WL_winding}.
We then know that we can only have $x$ being odd, since an even $x$ will make $\dsZ_2 = 0 $.
Now suppose there is a gap between $\phi_{l_0}$ and $\phi_{l_0+1}$.
$l_0$ cannot be odd, since odd $l_0$ will make $\phi_{l_0+1}(k_1)$ and $\phi_{l_0}(k_1)$ touching at $k_1 = 0$.
However, $l_0$ cannot be even either.
Because for each nontrivial Kramers pairs $e^{\ii \phi_{l_1} (\pi)}$ and $e^{\ii \phi_{l_2} (\pi)}$ with $l_1< l_2$, we must have $l_1 \leq l_0$ and $l_2 \geq l_0+1$, which means $(l_0 - x)$ WL bands with $l\leq l_0$ must all form trivial Kramers pairs to realize the gap and thus $(l_0 -x) $ is even.
However, if $l_0$ is even, then $(l_0 - x)$ must be odd, leading to contradiction.
Therefore, $l_0$ cannot be even either, which means $l_0$ does not exist, \ie, it is impossible to have a direct gap among $\phi_{l}(k_1)$'s

The connectivity (\ie, the absence of direct gap) and the no-crossing condition (\cref{eq:no_crossing_conditions}) tell us that $\phi_l(0) = \phi_{l+1}(0)$ for odd $l$ and $\phi_{l}(\pi) = \phi_{l+1}(\pi)$ for even $l$.
Thus, 
\eqa{
& \sum_l \left| \int_0^\pi dk_1 \frac{d}{d k_1} \phi_l(k_1) \right| \geq  \left[ \sum_{l=1,3,5,...,2N-1}  \phi_l(0) - \phi_l(\pi) \right] + \left[ \sum_{l=2,4,6,...,2N}  \phi_l(\pi) - \phi_l(0) \right] \\
& = \sum_{l=1,3,5,...,2N-1} [   \phi_l(0) -  \phi_{l+1}(0) ]  + \sum_{l=2,4,6,...,2N-2} \left[   \phi_l(\pi) -  \phi_{l+1}(\pi) \right] + \phi_{2N}(\pi) - \phi_1(\pi) =  \phi_{2N}(\pi) - \phi_1(\pi)  = 2\pi n \geq  2 \pi.
}
\cref{eq:winding_Z2_bound} is proven for $\dsZ_2 = 1$.

Combining \cref{eq:winding_Z2_bound} with \cref{eq:WL_lower_bound_of_Trg_for_Z_2}, we arrive at 
\eq{
2\pi \dsZ_2 \leq \int_{\text{half BZ}}  d^2 k\ 2\sqrt{\det(g_{\bsl{k}})} \leq \int_{\text{half BZ}} d^2 k   \Tr[g_{\bsl{k}}]\ .
}
Combined with the fact that $g_{\bsl{k}}= g_{-\bsl{k}}$ owing to TR symmetry, we obtain
\eq{
\label{eq:g_Z2_bound}
\frac{1}{2\pi} \int_{\text{BZ}} d^2 k    \Tr[g_{\bsl{k}}]  \geq \frac{1}{2\pi} \int_{\text{BZ}} d^2 k\ 2\sqrt{\det(g_{\bsl{k}})}  \geq 2 \dsZ_2\ .
}
We note that the PH $\dsZ_2$ index provides the same bound as \cref{eq:g_Z2_bound}.

We also note that the bound for integrated quantum metric can be directly applied to quantum distance, since the infinitesimal change of quantum distance is the quantum metric~\cite{Rhim_2020_quantum_distance_flat_bands}.
Specifically, the quantum distance for an isolated set of $N$ bands can be defined as
\eq{
d_{\bsl{k},\bsl{k}'} = N - \Tr[P_{\bsl{k}}P_{\bsl{k}'}]\ .
}
Then, for $\bsl{k}'=\bsl{k}+d\bsl{k}$, we have (to the second order)
\eqa{
d_{\bsl{k},\bsl{k}+d\bsl{k}} & =  - \Tr[P_{\bsl{k}}(d\bsl{k}\cdot\nabla_{\bsl{k}})P_{\bsl{k}}] -  \frac{1}{2}\Tr[P_{\bsl{k}}(d\bsl{k}\cdot\nabla_{\bsl{k}})^2P_{\bsl{k}}] \\
& =  - d\bsl{k}\cdot\nabla_{\bsl{k}} \frac{1}{2}\Tr[P_{\bsl{k}}P_{\bsl{k}}] -  \frac{1}{2}d\bsl{k}\cdot\nabla_{\bsl{k}}\Tr[P_{\bsl{k}} (d\bsl{k}\cdot\nabla_{\bsl{k}}) P_{\bsl{k}}] +   \frac{1}{2}\Tr[(d\bsl{k}\cdot\nabla_{\bsl{k}}P_{\bsl{k}})(d\bsl{k}\cdot\nabla_{\bsl{k}}P_{\bsl{k}})] \\
& = \frac{1}{2}\Tr[(d\bsl{k}\cdot\nabla_{\bsl{k}}P_{\bsl{k}})(d\bsl{k}\cdot\nabla_{\bsl{k}}P_{\bsl{k}})] \\
& = \sum_{i,j=x,y}dk_i dk_j [g_{\bsl{k}}]_{ij}\ .
}
Therefore, we have
\eq{
\left. \frac{1}{2}\nabla_{\bsl{k}'}^2 d_{\bsl{k},\bsl{k}'}\right|_{\bsl{k}'\rightarrow \bsl{k}}  = \Tr[g_{\bsl{k}}]\ ,
}
whose integration is bounded from below.
One interesting future direction is to bound other properties of quantum distance.
We note that if we pick the larger diagonal element of the quantum metric, labeled as $g_1(\bsl{k})$, it is bounded from below by $\Tr[g_{\bsl{k}}]/2$, and thus the integrated $g_1(\bsl{k})$ is bounded from below by half of the lower bound of integrated $\Tr[g_{\bsl{k}}]/2$.
For the smaller diagonal element $g_2(\bsl{k})$, it is bounded from below $\det(g_{\bsl{k}})/g_{1,max}$ ($g_{1,max}$ is the maximum value of $g_1(\bsl{k})$), and thus the integrated $\sqrt{g_2(\bsl{k})}$ is bounded from below by $1/(2 g_{1,max})$ times the lower bound of integrated $2\sqrt{\det(g_{\bsl{k}})}$.
However, the same bounds do not exist for the off-diagonal element of the quantum metric, since they can be negative in general.

\section{Physical Consequences of the WL Bound}

In this section, we will discuss three physical consequences, superfluid weight, optical conductivity and band gap, that can be bounded by the WL bound in \cref{eq:g_WL_bound} and thus $\dsZ_2 $ bound in \cref{eq:g_Z2_bound}.
We always choose
\eq{
\hbar=e=1
}
with electron having charge $-e$.

\subsection{WL Lower Bound for Superfluid Weight}
\label{app:SW_bound}

We first discuss the mean-field superfluid weight in flat bands, which we show is bounded from below by the absolute WL winding (and thus the $\dsZ_2$ index) according to the inequality proved in the Main Text.
The derivation that relates superfluid weight to quantum geometry has been worked out in \refcite{Torma2015SWBoundChern,Torma2016SuperfluidWeightLieb,Liang2017SWBandGeo,Hu2019MATBGSW,Xie2020TopologyBoundSCTBG,Torma2020SFWTBG,Rossi2021CurrentOpinion,Yu2022EOCPTBG,Torma2023WhereCanQuantumGeometryLeadUs,Tian2023QuantumGeoSC} and is included here for completeness. 
The new result here is the new bound of superfuild weight due to the general absolute WL winding and the $\dsZ_2$ index.

We consider a normal-phase TR-invariant 2D single-particle Hamiltonian $H_0$ that can realize a exactly-flat (doubly degenerate) band at energy $E_0$. If the flat bands have nontrivial topology such as $\dsZ_2$, the Hamiltonian is always allowed in the continuous Hamiltonian (such as moiré Hamiltonians) but may have long-range hoppings in a tight-binding formalism.
In general, $H_0$ reads
\eq{
H_0 = \sum_{\bsl{k},i,j} c^\dagger_{\bsl{k},i} \left[ h_0(\bsl{k})  \right]_{ij} c_{\bsl{k},j} \ , 
}
Suppose when the flat band is partially filled, we will have superconductivity with order parameter
\eq{
H_{\Delta} = \frac{1}{2} \Delta \sum_{\bsl{k}} c^\dagger_{\bsl{k} }U_{\TR}  \left( c_{-\bsl{k}}^\dagger \right)^T + h.c. \ ,
}
where 
\eq{
c^\dagger_{\bsl{k} } = ( \dots, c^\dagger_{\bsl{k},i},\dots )\ ,
}
and $U_{\TR}$ is the matrix representation of the TR symmetry, \ie,
\eq{
\TR c^\dagger_{\bsl{k}} \TR^{-1} = c^\dagger_{-\bsl{k}} U_{\TR} \ ,
}
where $U_{\TR}$ is a unitary matrix that satisfies
\eq{
U_{\TR} U_{\TR}^* = - 1 \Rightarrow U_{\TR}  = - U_{\TR}^T\ . 
}
Then, the mean-field Hamiltonian reads
\eq{
H_{MF} = \sum_{\bsl{k}} c^\dagger_{\bsl{k}} \left[ h_0(\bsl{k}) -\mu \right] c_{\bsl{k}} +  \frac{1}{2} \Delta \sum_{\bsl{k}} c^\dagger_{\bsl{k} }U_{\TR}  \left( c_{-\bsl{k}}^\dagger \right)^T +\frac{1}{2} \Delta^* \sum_{\bsl{k}} c^T_{-\bsl{k} }U_{\TR}^\dagger   c_{\bsl{k}}  \ .
}
For examples of microscopic Hamiltonians where the above mean-field description of uniform pairing described by a single number $\Delta$ is accurate, we refer to reader to \cite{tovmasyan2016effective,Herzog2022ManyBodySCFlatBand}. 

We now focus on the superfluid weight \cite{Torma2022ReviewQuantumGeometry}.
For that purpose, we will add thread flux in the mean-field Hamiltonian, and obtain 
\eq{
H_{MF}(\bsl{A}) = \sum_{\bsl{k}} c^\dagger_{\bsl{k}} \left[ h_0(\bsl{k}+\bsl{A}) -\mu \right] c_{\bsl{k}} +  \frac{1}{2} \Delta \sum_{\bsl{k}} c^\dagger_{\bsl{k} }U_{\TR}  \left( c_{-\bsl{k}}^\dagger \right)^T +\frac{1}{2} \Delta^* \sum_{\bsl{k}} c^T_{-\bsl{k} }U_{\TR}^\dagger   c_{\bsl{k}}  \ .
}
Let us consider the case that the flat $\dsZ_2$ band is well isolated, and thus we can project the mean-field Hamiltonian to that specific band.
To do so, we use $U_{n,\bsl{k}}$ to label the eigenvectors for the flat $\dsZ_2$ band ($n=1,2$).
The creation operator for the $\dsZ_2$ band reads
\eq{
\psi^{\dagger}_{n,\bsl{k},\bsl{A}} = c^\dagger_{\bsl{k}} U_{n,\bsl{k}+\bsl{A}}\ ,
}
and thus the projected mean-field Hamiltonian reads
\eqa{
\overline{H}_{MF}(\bsl{A}) & = \sum_{\bsl{k}} \psi^\dagger_{\bsl{k},\bsl{A}} \left[ E_0 -\mu \right] \psi_{\bsl{k},\bsl{A}} +  \frac{1}{2}  \sum_{\bsl{k}} \psi^\dagger_{\bsl{k} ,\bsl{A}} \Delta_{\bsl{k},\bsl{A}} \left( \psi_{-\bsl{k},\bsl{A}}^\dagger \right)^T +\frac{1}{2}  \sum_{\bsl{k}} \psi^T_{-\bsl{k} ,\bsl{A}} \Delta_{\bsl{k},\bsl{A}}^\dagger  \psi_{\bsl{k} ,\bsl{A}} \\
& = \frac{1}{2} \sum_{\bsl{k}} \mat{ \psi^\dagger_{\bsl{k},\bsl{A}} & \psi_{-\bsl{k},\bsl{A}}^T   } h_{\text{BdG}}(\bsl{k},\bsl{A})
\mat{ \psi_{\bsl{k} ,\bsl{A}} \\ \left(\psi^\dagger_{-\bsl{k},\bsl{A}} \right)^T }.
}
where $\psi^{\dagger}_{\bsl{k},\bsl{A}}=(\psi^{\dagger}_{1,\bsl{k},\bsl{A}},\psi^{\dagger}_{2,\bsl{k},\bsl{A}})$, 
\eq{
\label{eq:Delta}
\Delta_{\bsl{k},\bsl{A}} = \Delta U_{\bsl{k}+\bsl{A}}^\dagger U_{\TR}  U_{-\bsl{k}+\bsl{A}}^*\ ,
}
\eq{
h_{\text{BdG}}(\bsl{k},\bsl{A}) = \mat{ \left[ E_0 -\mu \right] \sigma_0 & \Delta_{\bsl{k},\bsl{A}} \\ 
\Delta_{\bsl{k},\bsl{A}}^\dagger & -  \left[ E_0 -\mu \right] \sigma_0 }\ ,
}
and
\eq{
U_{\bsl{k}} = \mat{ U_{1,\bsl{k}} &  U_{2,\bsl{k}}}\ .
}
The order parameter in \eqnref{eq:Delta} naturally obeys the requirement from the fermion statistics:
\eq{
\label{eq:Delta_transport}
\Delta_{\bsl{k},\bsl{A}}^T = \Delta    U_{-\bsl{k}+\bsl{A}}^\dagger U_{\TR}^T U_{\bsl{k}+\bsl{A}}^* = \Delta_{-\bsl{k},\bsl{A}} \ .
}
The TR symmetry provides
\eq{
U_{\TR}  U_{\bsl{k}}^* = U_{-\bsl{k}} D_{\TR}(\bsl{k})
}
with unitary $D_{\TR}(\bsl{k})$ satisfying
\eq{
D_{\TR}(-\bsl{k}) D_{\TR}^*(\bsl{k}) = -1\ .
}
As a result, we have
\eq{
\Delta_{\bsl{k},\bsl{A}} = \Delta U_{\bsl{k}+\bsl{A}}^\dagger U_{\TR}  U_{-\bsl{k}+\bsl{A}}^* = \Delta U_{\bsl{k}+\bsl{A}}^\dagger U_{\bsl{k}-\bsl{A}}  D_{\TR}(-\bsl{k}+\bsl{A})
}

The superfluid weight is derived from the free energy, which reads
\eqa{
\Omega(\bsl{A}) &  = \sum_{\bsl{k}}\left( \frac{1}{2} \Tr[ ( E_0 -\mu ) \sigma_0] - \frac{1}{\beta} \sum_{n}^{E_{\bsl{k},n}(\bsl{A})>0} \log\left( 2 \cosh(\beta E_{\bsl{k},n}(\bsl{A})/2) \right) \right) \\
 & = \sum_{\bsl{k}}\left(  ( E_0 -\mu )  - \frac{1}{\beta} \sum_{n}^{E_{\bsl{k},n}(\bsl{A})>0} \log\left( 2 \cosh(\beta E_{\bsl{k},n}(\bsl{A})/2) \right) \right)\ ,
}
where $E_{\bsl{k},n}(\bsl{A})$ is the eigenvalue of $h_{\text{BdG}}(\bsl{k},\bsl{A})$.
To derive the analytical expression for $E_{\bsl{k},n}(\bsl{A})$, let us consider $h_{\text{BdG}}^2(\bsl{k},\bsl{A})$, which reads
\eq{
h_{\text{BdG}}^2(\bsl{k},\bsl{A})  =\left[ E_0 -\mu \right]^2+  \mat{  \Delta_{\bsl{k},\bsl{A}} \Delta_{\bsl{k},\bsl{A}}^\dagger &   \\ 
  &  \Delta_{\bsl{k},\bsl{A}}^\dagger \Delta_{\bsl{k},\bsl{A}} } \ .
}
The TR symmetry provides
\eqa{
\label{eq:DeltaDeltaDag}
\Delta_{\bsl{k},\bsl{A}} \Delta_{\bsl{k},\bsl{A}}^\dagger & = \left|\Delta \right|^2 U_{\bsl{k}+\bsl{A}}^\dagger U_{\bsl{k}-\bsl{A}}  D_{\TR}(-\bsl{k}+\bsl{A}) D_{\TR}^\dagger (-\bsl{k}+\bsl{A}) U_{\bsl{k}-\bsl{A}}^\dagger  U_{\bsl{k}+\bsl{A}} \\
& = \left|\Delta \right|^2 U_{\bsl{k}+\bsl{A}}^\dagger U_{\bsl{k}-\bsl{A}} U_{\bsl{k}-\bsl{A}}^\dagger  U_{\bsl{k}+\bsl{A}} \ .
}
Combining the TR symmetry with \eqnref{eq:Delta_transport}, we arrive at 
\eqa{
& D_{\TR}^\dagger(-\bsl{k}-\bsl{A})  \Delta_{\bsl{k},\bsl{A}} \Delta_{\bsl{k},\bsl{A}}^\dagger  D_{\TR} (-\bsl{k}-\bsl{A}) \\
& = \left|\Delta \right|^2 D_{\TR}^\dagger(-\bsl{k}-\bsl{A}) U_{\bsl{k}+\bsl{A}}^\dagger U_{\bsl{k}-\bsl{A}} U_{\bsl{k}-\bsl{A}}^\dagger  U_{\bsl{k}+\bsl{A}} D_{\TR} (-\bsl{k}-\bsl{A}) \\
& =  \left|\Delta \right|^2 U_{-\bsl{k}-\bsl{A}}^T U_{\TR}^\dagger U_{\TR} U_{-\bsl{k}+\bsl{A}}^* U_{-\bsl{k}+\bsl{A}}^T   U_{\TR}^\dagger U_{\TR} U_{-\bsl{k}-\bsl{A}}^* \\
& = \left|\Delta \right|^2 U_{-\bsl{k}-\bsl{A}}^T U_{-\bsl{k}+\bsl{A}}^* U_{-\bsl{k}+\bsl{A}}^T   U_{-\bsl{k}-\bsl{A}}^* \\
&=  D_{\TR}^* (\bsl{k}+\bsl{A}) \left[ \Delta_{-\bsl{k},\bsl{A}}^\dagger  \Delta_{-\bsl{k},\bsl{A}} \right]^T D_{\TR}^T(\bsl{k}+\bsl{A}) \\
& = D_{\TR}^* (\bsl{k}+\bsl{A})  \Delta_{\bsl{k},\bsl{A}}^\dagger  \Delta_{\bsl{k},\bsl{A}}  D_{\TR}^T(\bsl{k}+\bsl{A}) \ ,
}
Therefore, the eigenvalues of $\Delta_{\bsl{k},\bsl{A}} \Delta_{\bsl{k},\bsl{A}}^\dagger$ are equal to those of $\Delta_{\bsl{k},\bsl{A}}^\dagger \Delta_{\bsl{k},\bsl{A}}$, which we labeled as $\lambda_{0,\bsl{k},\bsl{A}}$ and $\lambda_{1,\bsl{k},\bsl{A}}$.
Furthermore, based on the TR symmetry, we have
\eqa{
\Delta_{\bsl{k},\bsl{A}}^\dagger  \Delta_{\bsl{k},\bsl{A}} & = \left|\Delta \right|^2 D_{\TR}^\dagger (-\bsl{k}+\bsl{A}) U_{\bsl{k}-\bsl{A}}^\dagger  U_{\bsl{k}+\bsl{A}} U_{\bsl{k}+\bsl{A}}^\dagger U_{\bsl{k}-\bsl{A}}  D_{\TR}(-\bsl{k}+\bsl{A}) \\
& = D_{\TR}^\dagger (-\bsl{k}+\bsl{A})\Delta_{\bsl{k},-\bsl{A}} \Delta_{\bsl{k},-\bsl{A}}^\dagger  D_{\TR}(-\bsl{k}+\bsl{A})\ ,
}
which means we can always choose (and we will choose) $\lambda_{\alpha,\bsl{k},\bsl{A}}=\lambda_{\alpha,\bsl{k},-\bsl{A}}$ for $\alpha=0,1$.
The eigenvalues of $h_{\text{BdG}}(\bsl{k},\bsl{A})$ are just 
\eq{
E_{s,\alpha}(\bsl{k}) = (-1)^s \sqrt{ (E_0-\mu)^2 + \lambda_{\alpha,\bsl{k},\bsl{A}} }
}
for $s=0,1$ and $\alpha=0,1$.
Clearly, the positive $E_{+,\alpha}(\bsl{k})$ are non-negative, which gives
\eqa{
\Omega(\bsl{A}) 
 & = \sum_{\bsl{k}}\left(  ( E_0 -\mu )  - \frac{1}{\beta} \sum_{\alpha} \log\left( 2 \cosh(\frac{\beta}{2} \sqrt{ (E_0-\mu)^2 + \lambda_{\alpha,\bsl{k},\bsl{A}} }) \right) \right)\ .
}
The superfluid weight just reads
\eq{
\left[D_s(T)\right]_{ij} = \frac{1}{\mathcal{V}} \left. \frac{\partial^2 \Omega(\bsl{A}) }{\partial A_i \partial A_j}  \right|_{\bsl{A}\rightarrow 0}\ ,
}
where $\mathcal{V}$ is the sample volume. In a fully self-consistent treatment, the partial derivatives must be replaced with total derivatives. Nevertheless, it was shown in \cite{Huhtinen2022FlatBandSCQuantumMetric} that TR symmetry guarantees the same result holds as long as $g_{\bm{k}}$ is the minimal quantum metric. 

We are particularly interested in the zero-temperature case, where the free energy becomes
\eq{
\label{eq:free_energy_zero_temperature}
\Omega_0(\bsl{A})  = \sum_{\bsl{k}}\left(  ( E_0 -\mu )  - \frac{1}{2}  \sum_{\alpha} \sqrt{ (E_0-\mu)^2 + \lambda_{\alpha,\bsl{k},\bsl{A}} } \right)\ .
}
As a result, the zero-temperature superfluid weight reads
\eqa{
 \mathcal{V}\left[D_{SW}(0)\right]_{ij} & = \left. \frac{\partial^2 }{\partial A_i \partial A_j} \sum_{\bsl{k}}\left(  ( E_0 -\mu )  - \frac{1}{2}  \sum_{\alpha} \sqrt{ (E_0-\mu)^2 + \lambda_{\alpha,\bsl{k},\bsl{A}} } \right) \right|_{\bsl{A}\rightarrow 0} \\
& = - \frac{1}{2} \sum_{\bsl{k}}\sum_{\alpha}  \left. \frac{\partial^2 }{\partial A_i \partial A_j}  \sqrt{ (E_0-\mu)^2 + \lambda_{\alpha,\bsl{k},\bsl{A}} }  \right|_{\bsl{A}\rightarrow 0} \\
& = - \frac{1}{2} \sum_{\bsl{k}}\sum_{\alpha}  \left. \frac{\partial }{\partial A_i }  \frac{1}{2 \sqrt{ (E_0-\mu)^2 + \lambda_{\alpha,\bsl{k},\bsl{A}} }} \frac{\partial}{\partial A_j}\lambda_{\alpha,\bsl{k},\bsl{A}}  \right|_{\bsl{A}\rightarrow 0} \\
& = - \frac{1}{2} \sum_{\bsl{k}}\sum_{\alpha}  \left[ -\frac{1}{4 [(E_0-\mu)^2 + \lambda_{\alpha,\bsl{k},\bsl{A}} ]^{3/2}} \frac{\partial \lambda_{\alpha,\bsl{k},\bsl{A}}   }{\partial A_i } \frac{\partial \lambda_{\alpha,\bsl{k},\bsl{A}}  }{\partial A_j} +  \frac{1}{2 \sqrt{ (E_0-\mu)^2 + \lambda_{\alpha,\bsl{k},\bsl{A}} }} \frac{\partial^2}{\partial A_i  \partial A_j}\lambda_{\alpha,\bsl{k},\bsl{A}}  \right]_{\bsl{A}\rightarrow 0}\ .
}
From \eqnref{eq:DeltaDeltaDag}, we know 
\eqa{
\mathcal{V} \Delta_{\bsl{k},0} \Delta_{\bsl{k},0}^\dagger = \left|\Delta \right|^2  \sigma_0 \ ,
}
which means
\eq{
\lambda_{0,\bsl{k},0}=\lambda_{1,\bsl{k},0}=\left|\Delta \right|^2\ .
}
Therefore, the zero-temperature superfluid weight becomes 
\eqa{
 \mathcal{V} \left[D_{SW}(0)\right]_{ij}
& = - \frac{1}{2} \sum_{\bsl{k}}\sum_{\alpha}  \left[ -\frac{1}{4 [(E_0-\mu)^2 + \left|\Delta \right|^2 ]^{3/2}} \frac{\partial \lambda_{\alpha,\bsl{k},\bsl{A}}   }{\partial A_i } \frac{\partial \lambda_{\alpha,\bsl{k},\bsl{A}}  }{\partial A_j} +  \frac{1}{2 \sqrt{ (E_0-\mu)^2 +\left|\Delta \right|^2 }} \frac{\partial^2}{\partial A_i  \partial A_j}\lambda_{\alpha,\bsl{k},\bsl{A}}  \right]_{\bsl{A}\rightarrow 0}\\
& = - \frac{1}{2} \sum_{\bsl{k}}\sum_{\alpha}  \left[ -\frac{1}{4 [(E_0-\mu)^2 + \left|\Delta \right|^2 ]^{3/2}} \frac{\partial \lambda_{\alpha,\bsl{k},\bsl{A}}   }{\partial A_i } \frac{\partial \lambda_{\alpha,\bsl{k},\bsl{A}}  }{\partial A_j} +  \frac{1}{2 \sqrt{ (E_0-\mu)^2 +\left|\Delta \right|^2 }} \frac{\partial^2}{\partial A_i  \partial A_j}\lambda_{\alpha,\bsl{k},\bsl{A}}  \right]_{\bsl{A}\rightarrow 0}\ .
}
Since $\lambda_{\alpha,\bsl{k},\bsl{A}}=\lambda_{\alpha,\bsl{k},-\bsl{A}}$, we have
\eq{
\left. \frac{\partial \lambda_{\alpha,\bsl{k},\bsl{A}}   }{\partial A_i }\right|_{\bsl{A}\rightarrow 0 } =  0\ .
}
Then, 
\eqa{
  \mathcal{V} \left[D_{SW}(0)\right]_{ij}
& = - \frac{1}{4 \sqrt{ (E_0-\mu)^2 +\left|\Delta \right|^2 }} \sum_{\bsl{k}}\sum_{\alpha}  \left[ \frac{\partial^2}{\partial A_i  \partial A_j}\lambda_{\alpha,\bsl{k},\bsl{A}}  \right]_{\bsl{A}\rightarrow 0}\\
& = - \frac{1}{4 \sqrt{ (E_0-\mu)^2 +\left|\Delta \right|^2 }} \sum_{\bsl{k}}  \left[ \frac{\partial^2}{\partial A_i  \partial A_j}\Tr[\Delta_{\bsl{k},\bsl{A}} \Delta_{\bsl{k},\bsl{A}}^\dagger]  \right]_{\bsl{A}\rightarrow 0}\\
& = - \frac{\left|\Delta \right|^2 }{4 \sqrt{ (E_0-\mu)^2 +\left|\Delta \right|^2 }} \sum_{\bsl{k}}  \left[ \frac{\partial^2}{\partial A_i  \partial A_j}\Tr[ P_{\bsl{k}-\bsl{A}} P_{\bsl{k}+\bsl{A}} ]  \right]_{\bsl{A}\rightarrow 0}\\
& = - \frac{\left|\Delta \right|^2 }{4 \sqrt{ (E_0-\mu)^2 +\left|\Delta \right|^2 }} \sum_{\bsl{k}}  \left[ \frac{\partial^2}{\partial A_i  \partial A_j}\Tr[ P_{\bsl{k}-\bsl{A}} P_{\bsl{k}+\bsl{A}} ]  \right]_{\bsl{A}\rightarrow 0}\\
& = - \frac{\left|\Delta \right|^2 }{4 \sqrt{ (E_0-\mu)^2 +\left|\Delta \right|^2 }} \sum_{\bsl{k}}  \left\{ \Tr[ \partial_{k_i}\partial_{k_j} P_{\bsl{k}} P_{\bsl{k}} ]  -  \Tr[ \partial_{k_i} P_{\bsl{k}} \partial_{k_j} P_{\bsl{k}} ]  - \Tr[ \partial_{k_j} P_{\bsl{k}}\partial_{k_i} P_{\bsl{k}} ]  + \Tr[  P_{\bsl{k}} \partial_{k_i}\partial_{k_j} P_{\bsl{k}} ]  \right\}\\
& =  \frac{\left|\Delta \right|^2 }{ \sqrt{ (E_0-\mu)^2 +\left|\Delta \right|^2 }} \sum_{\bsl{k}}  \Tr[ \partial_{k_i} P_{\bsl{k}} \partial_{k_j} P_{\bsl{k}} ] =  \frac{2\left|\Delta \right|^2 }{ \sqrt{ (E_0-\mu)^2 +\left|\Delta \right|^2 }} \sum_{\bsl{k}} [ g_{\bsl{k}}]_{ij}\\
}
where
\eq{
P_{\bsl{k}} = U_{\bsl{k}} U_{\bsl{k}}^\dagger \ ,
}
the second last equality uses
\eq{
\Tr[ \partial_{k_i}\partial_{k_j} P_{\bsl{k}} P_{\bsl{k}} ] = \partial_{k_i} \left( \Tr[ \partial_{k_j} P_{\bsl{k}} P_{\bsl{k}} ] \right) - \Tr[ \partial_{k_j} P_{\bsl{k}} \partial_{k_i} P_{\bsl{k}} ]  =  - \Tr[ \partial_{k_j} P_{\bsl{k}} \partial_{k_i} P_{\bsl{k}} ]
}
which comes from
\eq{
\Tr[ \partial_{k_j} P_{\bsl{k}} P_{\bsl{k}} ] = \frac{1}{2}\left( \Tr[ \partial_{k_j} P_{\bsl{k}} P_{\bsl{k}} ] + \Tr[ P_{\bsl{k}}\partial_{k_j}  P_{\bsl{k}} ] \right) =\frac{1}{2}\partial_{k_j}  \Tr[ P_{\bsl{k}} P_{\bsl{k}} ]  = 0\ ,
}
and
\eq{
[ g_{\bsl{k}}]_{ij} = \frac{1}{2} \Tr[ \partial_{k_i} P_{\bsl{k}} \partial_{k_j} P_{\bsl{k}} ]\ .
}
From the expression of \eqnref{eq:free_energy_zero_temperature}, we can obtain the zero-temperature average value of the electron number, which reads
\eq{
\left\langle N_e \right\rangle = - \partial_{i}\Omega_0(0)  = -\partial_{i}  \sum_{\bsl{k}}\left(  ( E_0 -\mu )  - \frac{1}{2}  \sum_{\alpha} \sqrt{ (E_0-\mu)^2 + |\Delta|^2} \right) =   N \left(  1 -  \frac{E_0-i}{\sqrt{ (E_0-\mu)^2 + |\Delta|^2}}  \right)\ ,
}
which means the filled portion of the normal-state bands is 
\eq{
f = \left\langle N_e \right\rangle/(2N) =  \frac{1}{2}\left[ 1- \frac{E_0-i}{\sqrt{ (E_0-\mu)^2 + |\Delta|^2}} \right]\ .
}
As 
\eq{
f(1-f) = \frac{1}{4}-(\frac{1}{2}-f)^2 =\frac{1}{4} - \frac{1}{4}\frac{(E_0-\mu)^2}{ (E_0-\mu)^2 + |\Delta|^2} = \frac{1}{4} \frac{|\Delta|^2 }{ (E_0-\mu)^2 + |\Delta|^2}\ ,
}
we arrive at 
\eq{
\left[D_{SW}(0)\right]_{ij} = 4  |\Delta|\sqrt{f(1-f)} \int \frac{d^2 k}{(2 \pi)^2}[ g_{\bsl{k}}]_{ij}
}
Thus, at the mean-field level, the zero-temperature superfluid weight is bounded from below by the $\dsZ_2$ index, \ie
\eq{
\Tr[D_s(0)] =4 |\Delta|\sqrt{f(1-f)} \int \frac{d^2 k}{(2 \pi)^2} \text{Tr}[g_{\bsl{k}}]  \geq 4 |\Delta|\sqrt{f(1-f)} \frac{ \overline{\N}}{4\pi^2} \ ,
}
where $\overline{\N}$ is the absolute WL winding for any proper deformation that ends up covering BZ.
Owing to \eqnref{eq:winding_Z2_bound}, we have for the $\dsZ_2$ invariant 
\eq{
\Tr[D_s(0)] =4 |\Delta|\sqrt{f(1-f)} \int \frac{d^2 k}{(2 \pi)^2} \text{Tr}[g_{\bsl{k}}] \geq \frac{4 |\Delta|}{\pi}  \sqrt{f (1-f)} \frac{\dsZ_2}{\pi} \ .
}
\subsection{WL Bounds for Optical Conductivity and Band Gap}
\label{app:Optical_bound}

In this part, we discuss how the WL (and thus $\dsZ_2$) bound affects the optical conductivity.
We will consider a 2D non-interacting band insulator with TR symmetry.
The optical conductivity tensor reads~\cite{Kubo1957,SWM2000}
\eq{
\sigma_{ij}(\omega + \ii 0^+) = - \ii \int \frac{d^2 k}{(2\pi)^2}  \sum_{ m,n }  \frac{ (\varepsilon_{n,\bsl{k}} - \varepsilon_{m,\bsl{k}})  [A_{\bsl{k},i}]_{n m}   [A_{\bsl{k},j}]_{m n}     }{ \omega + \varepsilon_{n,\bsl{k}} - \varepsilon_{m,\bsl{k}} + \ii 0^+}       (f_{n}-f_{m}) \ ,
}
where $0^+$ is an infinitesimal positive real number, $\varepsilon_{n,\bsl{k}}$ is the $n$th energy band, 
\eq{
f_n = \theta(\epsilon_n(\bsl{k}-\mu) )
}
with $i$ the chemical potential in the band gap, and
\eq{
[A_{\bsl{k},i}]_{n m}  = \ii \bra{u_{n,\bsl{k}}} \partial_{k_i} \ket{u_{m,\bsl{k}}}
}
with $\ket{u_{n,\bsl{k}}}$ the periodic part of the Bloch state.
It is known that~\cite{SWM2000}
\eq{
\int_0^\infty d\omega \frac{\Re[\sigma_{ii}(\omega + \ii 0^+)]}{\omega} =    \frac{1}{4\pi} \int d^2k\ g_{ii}(\bsl{k}) \ ,
}
where $g_{ij}(\bsl{k})$ is the quantum metric for the occupied band, and we have use the insulating property of the system.
Combined with our $\dsZ_2$ bound for quantum geometry (\cref{eq:g_Z2_bound}), we have a $\dsZ_2$ lower bound for the optical conductivity:
\eq{
\int_0^\infty \frac{d\omega}{\omega} \sum_i \Re[ \sigma_{ii}(\omega + \ii 0^+)] =    \frac{1}{4\pi} \int d^2k\ \Tr[g(\bsl{k})] \geq \frac{\overline{\N}}{4\pi} \ .
}
where $\overline{\N}$ is the absolute WL winding for any proper deformation that ends up covering BZ.
Owing to \eqnref{eq:winding_Z2_bound}, we have for the $\dsZ_2$ invariant
\eq{
\int_0^\infty \frac{d\omega}{\omega} \sum_i \Re[ \sigma_{ii}(\omega + \ii 0^+)] =    \frac{1}{4\pi} \int d^2k\ \Tr[g(\bsl{k})] \geq \dsZ_2\ .
}

In general, the non-interacting Hamiltonian for solids have the following form
\eq{
H = \int d^d r \sum_{s,s'} c^\dagger_{\bsl{r},s} \left[ -\frac{\bsl{\nabla}^2 }{2m} \delta_{ss'} -\ii \nabla\cdot \bsl{Y}_{ss'}(\bsl{r})  -\ii \bsl{Y}_{ss'}(\bsl{r}) \cdot \nabla  + V(\bsl{r}) \delta_{ss'}  \right] c_{\bsl{r},s'} \ ,
}
where $\bsl{r}$ is the position, $s$ is the spin, and 
\eq{
\left\{ c^\dagger_{\bsl{r},s} , c_{\bsl{r}',s'} \right\} = \delta(\bsl{r}-\bsl{r}') \delta_{ss'}\ .
}
$\bsl{Y}_{ss'}$ accounts for spin-orbit coupling.
The form of the Hamiltonian implies a sum rule for the optical conductivity, which reads~\cite{Kubo1957,Fu.Onishi.2023}
\eq{
\int_{0}^{\infty} d\omega \Re[\sigma_{ii}(-\omega + \ii 0^+)] = \frac{n \pi}{ m }\ ,
}
where $n$ is the electron density.
Combined with 
\eq{
\int_0^\infty \frac{d\omega}{\omega} \sum_i \Re[ \sigma_{ii}(\omega + \ii 0^+)] \leq \frac{1}{E_g} \int_0^\infty d\omega \sum_i \Re[ \sigma_{ii}(\omega + \ii 0^+)]
}
with $E_g$ the direct band gap, we arrive at a $\dsZ_2$ upper bound for the gap
\eq{
E_g \leq \frac{\int_0^\infty d\omega \sum_i \Re[ \sigma_{ii}(\omega + \ii 0^+)] }{\int_0^\infty \frac{d\omega}{\omega} \sum_i \Re[ \sigma_{ii}(\omega + \ii 0^+)] } = \frac{2n \pi}{ m } \frac{1}{\frac{1}{4\pi} \int d^2k\ \Tr[g(\bsl{k})]} \leq \frac{2n \pi}{ m } \frac{1}{\frac{\overline{\N}}{4\pi} }   \ ,
}
where $\overline{\N}$ is the absolute WL winding for any proper deformation that ends up covering BZ.
Owing to \eqnref{eq:winding_Z2_bound}, we have for the $\dsZ_2$ invariant
\eq{
E_g \leq \frac{2n \pi}{ m \dsZ_2} \ .
}
We note that the relation between the gap and the quantum metric was derived in \refcite{Kivelson1982,Fu.Onishi.2023}.

\end{document}